\documentclass[onecolumn,journal,draftclsnofoot,12pt]{IEEEtran}
\pdfoutput=1
\def\allfiles{}

\usepackage{bm}
\usepackage{amssymb}
\usepackage{amsmath}
\usepackage{amsfonts}
\usepackage{graphicx}
\usepackage{subfigure}
\usepackage{cite}
\usepackage{enumerate}
\usepackage{url}
\usepackage{epstopdf}
\usepackage{float}
\usepackage{color}
\usepackage{multirow}

\usepackage[normalem]{ulem}  
\newcommand{\beq}{\begin{equation}}
\newcommand{\eeq}{\end{equation}}
\newcommand{\blue}{\textcolor[rgb]{0,0,0.7}}
\newcommand{\pasted}{\textcolor[rgb]{0.3,0,0}}
\allowdisplaybreaks[4]
\makeatletter  
\newif\if@restonecol  
\makeatother  
\let\algorithm\relax  
\let\endalgorithm\relax  
\usepackage[linesnumbered, lined, commentsnumbered, ruled]{algorithm2e}
\usepackage{amsmath}  

\usepackage{makecell} 

\renewcommand{\Pr}{\mathrm{Pr}}
\newcommand{\diag}{\mathrm{diag}}
\newcommand{\rank}{\mathrm{rank}}
\newcommand{\tx}{\mathrm{Tx}}
\newcommand{\rx}{\mathrm{Rx}}
\newcommand{\rl}{\mathrm{rl}}
\newcommand{\mau}{\mathrm{mu}}
\newcommand{\sign}{\mathrm{sign}}

\DeclareMathOperator*{\argmin}{argmin}
\DeclareMathOperator*{\argmax}{argmax}

\newif\ifshowannote
\showannotetrue

\showannotefalse

\newif\ifBreakPage
\BreakPagefalse

\newif\ifShowIntro
\ShowIntrotrue

\newif\ifIntroduceDRL
\IntroduceDRLfalse

\newif\ifShowSimul
\ShowSimultrue

\newif\ifShowRes
\ShowRestrue

\newif\ifShowConclu
\ShowConclufalse

\newif\ifShowAppendix
\ShowAppendixtrue

\newcounter{optcnt}
\begin{document}

\title{\Large{Reconfigurable Intelligent Surfaces based RF Sensing: Design, Optimization, and Implementation}}

\author{
\IEEEauthorblockN{
\normalsize{Jingzhi~Hu},~\IEEEmembership{\normalsize Student~Member,~IEEE},
\normalsize{Hongliang~Zhang},~\IEEEmembership{\normalsize Member,~IEEE},
\normalsize{Boya~Di},~\IEEEmembership{\normalsize Member,~IEEE},
\normalsize{Lianlin~Li},~\IEEEmembership{\normalsize Senior~Member,~IEEE},
\normalsize{Lingyang~Song},~\IEEEmembership{\normalsize Fellow,~IEEE},
\normalsize{Yonghui~Li},~\IEEEmembership{\normalsize Fellow,~IEEE},
\normalsize{Zhu~Han},~\IEEEmembership{\normalsize Fellow,~IEEE},
and~\normalsize{H.~Vincent~Poor}~\IEEEmembership{\normalsize Fellow,~IEEE}
}
\thanks{
 J. Hu, L. Li, and L. Song are with Department of Electronics, Peking University.~(email: \{jingzhi.hu, lianlin.li, lingyang.song\}@pku.edu.cn)
 }
 \thanks{
H. Zhang is with Department of Electronics, Peking University, and also with Department of Electrical and Computer Engineering, University of Houston.~(email: hongliang.zhang92@gmail.com)
}
\thanks{B. Di is with Department of Electronics Engineering, Peking University, and also with Department of Computing, Imperial College London.~(email: diboya92@gmail.com)}
\thanks{
Y. Li is with School of Electrical and Information Engineering, University of Sydney.~(yonghui.li@sydney.edu.au)
}
\thanks{
Z. Han is with Electrical and Computer Engineering Department, University of Houston, and also with the Department of Computer Science and Engineering, Kyung Hee University.~(email: zhan2@uh.edu)
}
\thanks{
H. Vincent Poor is with Department of Electrical Engineering, Princeton University.~(poor@princeton.edu)
}
}

\maketitle
\vspace{-3.5em}
\begin{abstract}
\vspace{-1em}
Using radio-frequency~(RF) sensing techniques for human posture recognition has attracted growing interest due to its advantages of pervasiveness, contact-free observation, and privacy protection.
Conventional RF sensing techniques are constrained by their radio environments, which limit the number of transmission channels to carry multi-dimensional information about human postures.
Instead of passively adapting to the environment, in this paper, we design an RF sensing system for posture recognition based on reconfigurable intelligent surfaces~(RISs).
The proposed system can actively customize the environments to provide the desirable propagation properties and diverse transmission channels.
However, achieving high recognition accuracy requires the optimization of RIS configuration, which is a challenging problem.
To tackle this challenge, we formulate the optimization problem, decompose it into two subproblems and propose algorithms to solve them.
Based on the developed algorithms, we implement the system and carry out practical experiments.
Both simulation and experimental results verify the effectiveness of the designed algorithms and system. 
Compared to the random configuration and non-configurable environment cases, the designed system can greatly improve the recognition accuracy.
\end{abstract}
\vspace{-1em}
\begin{IEEEkeywords}
\vspace{-0.5em}
Reconfigurable intelligent surface, radio-frequency sensing, posture recognition.
\end{IEEEkeywords}
\newpage

\ifShowIntro
\ifx\allfiles\undefined
\documentclass[onecolumn,journal,draftclsnofoot,12pt]{IEEEtran}

\usepackage{bm}
\usepackage{amssymb}
\usepackage{amsmath}
\usepackage{amsfonts}
\usepackage{graphicx}
\usepackage{subfigure}
\usepackage{cite}
\usepackage{enumerate}
\usepackage{url}
\usepackage{epstopdf}
\usepackage{float}
\usepackage{color}
\usepackage{multirow}

\usepackage[normalem]{ulem}  
\newcommand{\beq}{\begin{equation}}
\newcommand{\eeq}{\end{equation}}
\newcommand{\blue}{\textcolor[rgb]{0,0,0.7}}
\newcommand{\pasted}{\textcolor[rgb]{0.3,0,0}}
\allowdisplaybreaks[4]
\makeatletter  
\newif\if@restonecol  
\makeatother  
\let\algorithm\relax  
\let\endalgorithm\relax  
\usepackage[linesnumbered, lined, commentsnumbered, ruled]{algorithm2e}
\usepackage{amsmath}  

\usepackage{makecell} 

\renewcommand{\Pr}{\mathrm{Pr}}
\newcommand{\diag}{\mathrm{diag}}
\newcommand{\rank}{\mathrm{rank}}
\newcommand{\tx}{\mathrm{Tx}}
\newcommand{\rx}{\mathrm{Rx}}
\newcommand{\rl}{\mathrm{rl}}
\newcommand{\mau}{\mathrm{mu}}
\newcommand{\sign}{\mathrm{sign}}

\DeclareMathOperator*{\argmin}{argmin}
\DeclareMathOperator*{\argmax}{argmax}

\newif\ifshowannote
\showannotetrue

\newtheorem{definition}{\bf Definition}
\newtheorem{proposition}{\bf Proposition}
\newtheorem{theorem}{\bf Theorem}
\newtheorem{lemma}{\bf Lemma}
\newtheorem{proof}{Proof}

\begin{document}
\fi
\section{Introduction}
 
 
Recently, leveraging widespread radio-frequency~(RF) signals for wireless sensing applications is of special interests.
Different from methods based on wearable devices or surveillance cameras, the RF sensing techniques need no contection with sensing targets and will raise no privacy concerns.~\cite{Kianoush2017Device}.
The basic principle behind RF sensing is that the influence of the sensing objectives on the wireless signal propagation can be potentially recognized by the receivers~\cite{Lee2009Wireless}. 
In RF sensing, posture recognition has been one of the most commonly studied topics with many applications such as surveillance~\cite{He2006Vigilnet}, ambient assisted living~\cite{cook2009assessing}, and remote health monitoring~\cite{amin2016radar}.
It is crucial to design RF sensing systems with high posture recognition accuracy.

RF posture recognition aims to automatically recognize different human postures such as standing, walking, sitting, and lying by analyzing the propagation characteristics and impacts of different postures on wireless signal propagation between sensors and receivers~\cite{Le2013Human}.
Such posture information can be extracted from the received signals.
In literature, several wireless posture recognition systems have been proposed.
In~\cite{Kellogg2014Bringing}, the authors designed a system to recognize different human postures through the signals extracted by an RFID tag.
In~\cite{Wang2016RF-Fall}, the human fall is detected by analyzing the phase shift of the received signals via a pair of Wi-Fi devices.
It is shown that the accuracy of posture recognition increases with the number of independent reflected paths between the transmitter and receiver, as multi-dimensional posture information are carried in the channels.
To increase the number of paths, the multi-antenna transceivers were employed to recognize postures such as standing, sitting, and lying~\cite{sasakawa2018human}.
In~\cite{adib2015capturing}, multiple transceivers were used to identify different parts of human body, which can be used to recognize human postures.

However, due to the complicated and unpredictable wireless environments, the accuracy and flexibility of posture recognition are greatly affected by unwanted multi-path fading~\cite{honma2018Human} and the limited number of independent channels from the transmitters to the receivers in the conventional RF sensing systems.
Recently, the reconfigurable intelligent surface~(RIS) technique has been developed as a promising technology to actively customize the propagation channels to create favorable propagation environment~\cite{Renzo2019Smart}.
By optimizing and programming the configurations, the RIS is able to customize the wireless channels and generate a favorable massive number of independent paths to enhance the posture recognition accuracy~\cite{zhou2017short}.

In this paper, we propose an RIS-based RF sensing system for human posture recognition\footnote{ 
In this paper, we focus on designing the RIS-based RF sensing systems for posture recognition.
Nevertheless, since the RISs have the capability of customizing radio environments for RF sensing, they can be adopted for various sensing applications, such as object detecting and localization systems, to enhance the performance.
}.
By periodically programming RIS configurations, the developed system can create multiple independent paths carrying out richer information of the human postures to achieve high accuracy of human posture recognitions. 

There are several challenges in designing an RIS-based posture recognition system.
\emph{First}, in order to obtain high posture recognition accuracy, RIS configurations need to be carefully designed to create the favorable propagation conditions for posture recognition at the receiver. 
However, the complexity of finding the optimal configuration is extremely high due to the large number of RIS elements and different states in each RIS element.
\emph{Second}, the decision function for posture recognition also greatly affects the recognition accuracy and is coupled with the RIS configuration optimizations. 
Therefore, it is necessary to jointly optimize the configuration and decision function to maximize recognition accuracy.

To tackle the above challenges, we decompose the problem into two subproblems, i.e., configuration optimization subproblem and decision function optimization subproblem, and then apply an alternating optimization algorithm and a supervised learning algorithm to solve these two subproblems, respectively.
More importantly, to demonstrate its benefits in practical systems, we implement our designed system using universal software radio peripheral~(USRP) devices and carry out simulations and practical experiments which verify the effectiveness of our system design and proposed algorithms.
The contributions of this paper can be summarized as below:
\begin{itemize}
    \item  We construct a human posture recognition system assisted by the cost-efficient RIS technique and USRP transceivers. By optimizing the RIS configuration to create the optimal propagation links, the system can accurately obtain information about the human postures.
    
    \item We formulate a joint RIS configuration and decision function optimization problem for the false recognition cost minimization. The formulated problem is decoupled and solved by our proposed algorithms, where the configuration and the decision function are optimized, respectively. The convergence of the proposed algorithms is analyzed.

	\item We implement the proposed design in practical testbed. Both simulations and experiments verify the effectiveness of the proposed system and algorithms. It is shown that the posture recognition accuracy increase with the size of the RIS and the number of independently controllable groups.  
		Compared with the random configuration and the non-configurable environment case in the conventional RF sensing systems, the RIS-based posture recognition with optimized configuration can significantly improve the recognition accuracy.
		\end{itemize}

The rest of this paper is organized as follows. In Section~\ref{sec:system}, we introduce the design of RIS based human posture recognition system. 
In Section~\ref{sec: problem formulation}, we formulate a problem to optimize the RIS configurations and the decision function to minimize the average cost of false posture reconfiguration, and develop algorithms to solve the problem in Section~\ref{sec: algorithm design}. 
In Section~\ref{sec: system implementation}, we introduce the system implementation.
Numerical and experimental results in Section~\ref{sec: results} validate our proposed algorithms and analysis. 
Finally, conclusions are drawn in Section~\ref{sec: conclusion}.

\ifx\allfiles\undefined
\end{document}
\fi
\fi
\ifBreakPage
\newpage
\fi

\ifShowIntro
\ifx\allfiles\undefined
\documentclass[onecolumn,journal,draftclsnofoot,12pt]{IEEEtran}

\usepackage{bm}
\usepackage{amssymb}
\usepackage{amsmath}
\usepackage{amsfonts}
\usepackage{graphicx}
\usepackage{subfigure}
\usepackage{cite}
\usepackage{enumerate}
\usepackage{url}
\usepackage{epstopdf}
\usepackage{float}
\usepackage{color}
\usepackage{multirow}

\usepackage[normalem]{ulem}  
\newcommand{\beq}{\begin{equation}}
\newcommand{\eeq}{\end{equation}}
\newcommand{\blue}{\textcolor[rgb]{0,0,0.7}}
\newcommand{\pasted}{\textcolor[rgb]{0.3,0,0}}
\allowdisplaybreaks[4]
\makeatletter  
\newif\if@restonecol  
\makeatother  
\let\algorithm\relax  
\let\endalgorithm\relax  
\usepackage[linesnumbered, lined, commentsnumbered, ruled]{algorithm2e}
\usepackage{amsmath}  

\usepackage{makecell} 

\renewcommand{\Pr}{\mathrm{Pr}}
\newcommand{\diag}{\mathrm{diag}}
\newcommand{\rank}{\mathrm{rank}}
\newcommand{\tx}{\mathrm{Tx}}
\newcommand{\rx}{\mathrm{Rx}}
\newcommand{\rl}{\mathrm{rl}}
\newcommand{\mau}{\mathrm{mu}}
\newcommand{\sign}{\mathrm{sign}}

\DeclareMathOperator*{\argmin}{argmin}
\DeclareMathOperator*{\argmax}{argmax}

\newif\ifshowannote
\showannotetrue

\newtheorem{definition}{\bf Definition}
\newtheorem{proposition}{\bf Proposition}
\newtheorem{theorem}{\bf Theorem}
\newtheorem{lemma}{\bf Lemma}
\newtheorem{proof}{Proof}

\begin{document}
\fi
\section{System Design}
\label{sec:system}

\begin{figure}[!t] 
	\center{\includegraphics[width=0.73\linewidth]{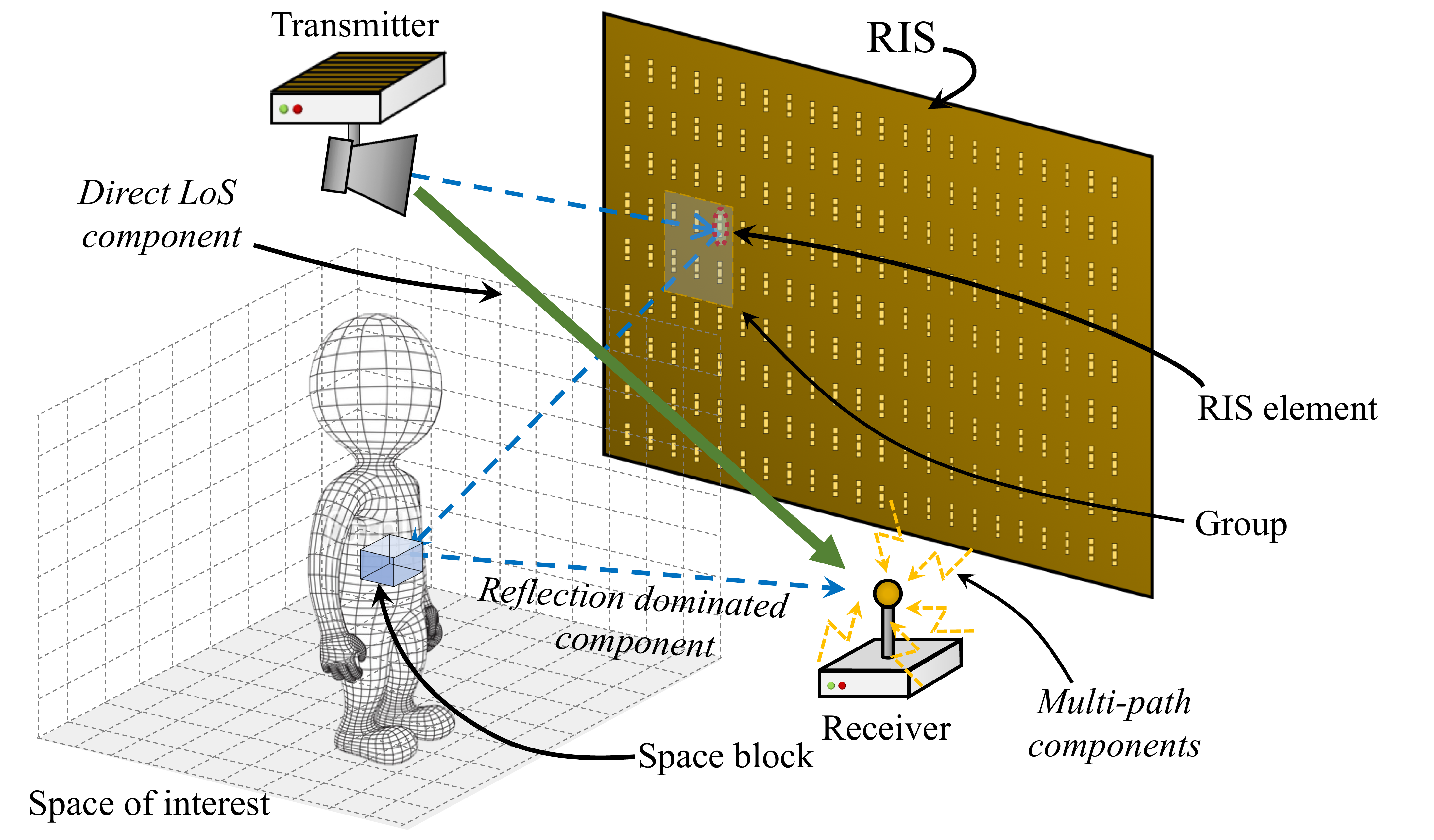}}
	\caption{Components of RIS-based posture recognition system.}
	\label{fig: channel_model}
\end{figure}

In this section, we propose the RIS-based posture recognition system.
As shown in Fig.~\ref{fig: channel_model}, the system is composed of a pair of single-antenna transceivers and an RIS.
The transmitter continuously transmits a single-tone RF signal of frequency $f_c$.
The RIS reflects the incident signals, which are then reflected by the human body and reach the receiver.

In the following, we first introduce the RIS in Section~\ref{ssec: ris model} and the channel model in Section~\ref{ssec: trans model}.
Then, we propose a periodic configuring protocol to perform posture recognition for the proposed system.

\subsection{RIS Model}
\label{ssec: ris model}
The RIS is an artificial thin film of electromagnetic and reconfigurable materials, which is composed of a large number of uniformly distributed and electrically controllable \emph{RIS elements}.
We denote the number of RIS elements by $N$ and the set of them by $\mathcal N$.
As shown in Fig.~\ref{fig: channel_model}, the RIS elements are arranged in a two-dimensional array.

Each RIS element is made of multiple metal patches connected by electrically controllable components, e.g., PIN diodes~\cite{Renzo2019Smart}, which are assembled on a dielectric surface.
Each PIN diode can be switched to either an \emph{ON} or \emph{OFF} state based on the applied bias voltages.
The \emph{state} of an RIS element is determined by the states of the PIN diodes on the RIS element.
Each state of the RIS element shows its own electrical property, leading to a unique reflection coefficient for the incident RF signals.
Suppose that an RIS element contains $N_D$ PIN diodes, and thus the RIS element can be configured into $2^{N_D}$ possible states.
We will describe the detailed implementation of the RIS elements in Section~\ref{sec: system implementation}. 

\begin{figure}[!t] 
	\center{\includegraphics[width=0.37\linewidth]{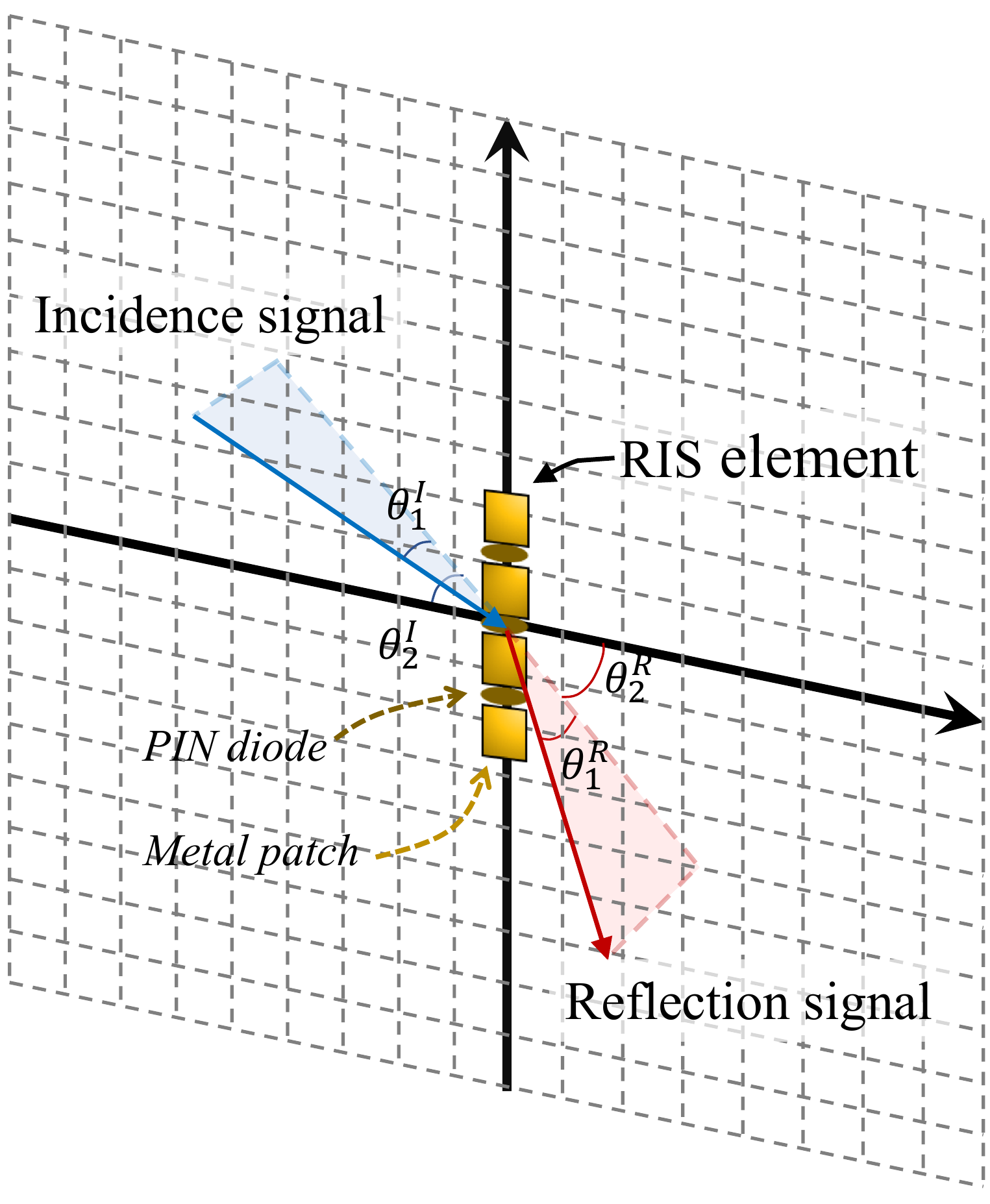}}
	\vspace{-1em}
	\caption{Incidence and reflection angles for signals reflected on an RIS element.}
	\label{fig: reflection coefficient}
\end{figure}
For simplicity, we refer to the set of all possible states of each RIS element as an \emph{available state set}, denoted by $\mathcal S_a$.
We denote that the number of available states by $N_a$, and the $i$-th~($i \in [1, N_a]$) state in $\mathcal S_a$ is denoted by $\hat{s}_i$.
For frequency $f_c$, the reflection coefficient of the RIS element can be expressed as $r(\bm \theta_I, \bm \theta_R, s)$, which is a function of the incidence angle $\bm \theta_I = (\theta_{I,1},\theta_{I,2})$, the reflection angle $\bm \theta_R=(\theta_{R,1}, \theta_{R,2})$, and state $s$.
An example of the incidence angle and the reflection angle is depicted in Fig.~\ref{fig: reflection coefficient}.
The value of reflection coefficient is a complex number, i.e., $r(\bm\theta^I, \bm \theta^R, s_n)\in\mathbb C$, and $|r(\bm\theta^I, \bm \theta^R, s)|$ and $\angle r(\bm\theta^I, \bm \theta^R, s)$ denote the amplitude ratio and the phase shift between the reflection and incidence signals, respectively.

However, since the number of RIS elements $N$ is usually large, it is costly and inefficient to control each RIS element independently.
To alleviate the controlling complexity, we divide the RIS elements into \emph{groups}.
Each group contains the same number of RIS elements, and the RIS elements in the same group are located within a square region, as shown in Fig.~\ref{fig: channel_model}. 
To be specific, the RIS elements are equally divided into $L$ groups, and the set of RIS elements in the $l$-th group is denoted $\mathcal N_l$, which satisfies $\mathcal N_l \cap \mathcal N_{l'} = \emptyset$~($\forall l,l'\in[1,L]$, $l\neq l'$) and $\bigcup_{l=1}^{L}\mathcal N_{l} = \mathcal N$. Moreover, we denote the size of group by $N_G = N/L$.

In the designed system, we control the RIS elements in the basic unit of the group, that is, the RIS elements in the same group are in the same state, and different groups of RIS elements are controlled independently.
We denote the state of the $l$-th group by $\nu_l$, i.e., $s_n=\nu_l$, $\forall n \in \mathcal N_l$.
We refer to the state vector of the $L$ groups, i.e., $\bm \nu$, as the \emph{configuration} of the RIS.
Through changing the configurations, the RIS is able to modify the waveforms of the reflected signals to form beamforming~\cite{di2019hybrid}.
By using the beamforming capability, the RIS is then able to generate various different waveforms which enhance posture recognition.

\subsection{Channel Model}
\label{ssec: trans model}
As shown in Fig.~\ref{fig: channel_model}, the pair of transceivers consists of a transmitter and a receiver, which are equipped with single antennas to transmit and receive RF signals, respectively.
The transmitter continuously transmits a unit baseband signal at carrier frequency $f_c$ with transmit power $P_t$.
The antenna at the transmitter is a directional antenna and is referred to as the \emph{Tx antenna}.
The main lobe of the Tx antenna is pointed towards the RIS, and therefore, most of the transmitted signals are reflected and modified by the RIS.
The modified signals enter the region in front of the RIS and are reflected by the human body at different positions.
The antenna of the receiver, referred to as the \emph{Rx antenna}, is an omni-directional vertical antenna, and thus all the signals reflected by the human body can be received.
Besides, the Rx antenna is located right below the RIS, so the signals reflected by the RIS are not received by the Rx antenna directly.

The transmission channel from the transmitter to the receiver can be modeled as a \emph{Rice channel}~\cite{zhang2019reconfigurable}, which is composed of  a \emph{direct line-of-sight~(LoS) component}, multiple \emph{reflection dominated components}, and a \emph{multi-path component}.
As shown in Fig.~\ref{fig: channel_model}, the direct LoS component accounts for signal path from the Tx antenna to the Rx antenna without any reflection; 
the reflection dominated component indicates the signals transmitted from the transmitter to the receiver via the shortest paths reflected from the RIS elements and human body.
The multi-path components account for the signals after the complex environment reflection and scattering.

To better describe the received signals at the receiver, we first define the region in front of the RIS as the \emph{space of interest}, where the human postures are positioned in.
Specifically, the space of interest is a $l_x \times l_y \times l_z$ cuboid region, as shown in Fig.~\ref{fig: channel_model}.
Besides, we discretize the space of interest into $M$ equally-sized \emph{space blocks}, as shown in Fig.~\ref{fig: channel_model}.
Based on \cite{Huang2006Effect}, the human body in the space of interest can be considered as a reflector for the wireless signals.
For generality, we denote the reflection coefficients of the $M$ space blocks as $\bm \eta = (\eta_1, ..., \eta_M)$, which is referred to as the \emph{space reflection vector}
\footnote{
For simplicity of description, we assume single human in the space of interest. 
Nevertheless, it can be observed in Section~\ref{sec: algorithm design} that the proposed algorithms are independent of the number of humans.
Therefore, the system and algorithm proposed in this paper can also be easily extended to the case of multi-objective posture recognition.
}.
Here, $\eta_m$~($m\in [1,M]$) is the reflection coefficient of the $m$-th space block for the signals reflected from the RIS to the receiver.
Intuitively, the space reflection vector is determined by the human postures in the space of interest.
For example, for a given posture, if a space block contains part of the human body, its reflection coefficient will be nonzero.
Otherwise, if the $m$-th space block is empty, $\eta_m = 0$.
In other words, the space reflection vector carries the information of human postures.

Given configuration $\bm \nu$ and space reflection vector $\bm \eta$, the received signal can be expressed as
\beq
\label{equ: received signal}
y= h_d\cdot P_t\cdot x + \sum_{m\in[1,M]}\sum_{l\in[1,L]}\sum_{n\in\mathcal N_l} h_{n,m}(\nu_l,\eta_m) \cdot P_t \cdot x + h_{\rl}\cdot P_t \cdot x +\sigma,
\eeq
where the first term indicates the signals of the direct LoS component, the second term is the signals of the $N\times M$ reflection dominated components, the third term represents the multi-path component, and $\delta$ denotes the noise signals.
Here, $h_d$ is the channel gain of the direct LoS path and can be calculated by
\beq
h_d = \frac{\lambda}{4\pi} \cdot \frac{\sqrt{g_{T,\mathrm{los}}g_{R,\mathrm{los}}}\cdot e^{-j2\pi d_{\mathrm{los}}/\lambda}}{d_{\mathrm{los}}},
\eeq
where $\lambda$ is the wavelength of the carrier signal, 
$g_{T,\mathrm{los}}$ and $g_{R,\mathrm{los}}$ denote the gains of the transmitter and the receiver for the direct LoS component, respectively,
and $d_{\mathrm{los}}$ is the distance of the direct LoS path from the Tx antenna to the Rx antenna.

Besides, $h_{n, m}(\nu_l, \eta_m)$ denotes the channel gain for the signals reflected by the $n$-th RIS element in the $l$-th group in state $\nu_l~(\nu_l\in\mathcal S_a)$ and by the $m$-th space block and then reach the receiver directly.
Based on~\cite{tang2019wireless}, channel gain $h_{n, m}(\nu_l, \eta_m)$ which involves an RIS element can be calculated by
\beq
\label{equ: main channel gain}
h_{n,m}(\nu_l, \eta_m) =\frac{\lambda\cdot r_{n,m}(\nu_l)\cdot\eta_{m}\cdot \sqrt{g_{T,n}g_{R,m}}\cdot e^{-j2\pi (d_n+d_{n,m})/\lambda}}{4\pi\cdot d_{n} \cdot d_{n,m}},
\eeq
where $r_{n,m}(\nu_l) = r(\bm \theta^{I}_n,\bm \theta^{R}_m, \nu_l)$ denotes the reflection coefficient of the $n$-th RIS element for the incidence signal towards the $m$-th space block in state $\nu_l$,
$g_{T,n}$ is the gain of the transmitter towards the $n$-th RIS element,
$g_{R,m}$ is the gain of the receiver towards the $m$-th space block,
$d_n$ is the distance from the Tx to the $n$-th RIS element,
and $d_{n,m}$ denotes the distance from the $n$-th RIS element to the Rx antenna via the $m$-th space block.
Finally, $h_{\rl}$ and $\sigma$ are random values following the complex normal distributions, i.e., which can be expressed as $h_{\rl}\sim \mathcal{CN}(0,\epsilon_{\rl})$, $\sigma\sim\mathcal{CN}(0,\epsilon_{n})$ with $\epsilon_{\rl}$ and $\epsilon_{n}$ being the variances.

\subsection{Protocol Design}
To coordinate the RIS and the transceiver in performing the posture recognition, we propose the \emph{periodic configuring protocol} as follows.
The timeline in the protocol is in unit of \emph{frame} with time duration $\delta$.
Moreover, instead for RIS to change the configurations in the unit of frame, in each frame, each RIS element changes from state $\hat{s}_1$ to $\hat{s}_{N_a}$ sequentially.
The duration that each RIS element is in each state is the parameter to be designed in frame configuration.
Under the proposed protocol, the limitations of the RIS due to the discreteness of the available state sets of RIS elements can be alleviated.
In a certain frame, for the $l$-th group, the time duration for the RIS elements to be in the $N_a$ available states are denoted by $\tilde{\bm t}_l = (\tilde{ t}_{l,1},...,\tilde{ t}_{l,N_a})^T$ with $\sum_{i = 1}^{N_a} \tilde{t}_{l,i} = \delta$.

\begin{figure}[!t] 
	\center{\includegraphics[width=0.7\linewidth]{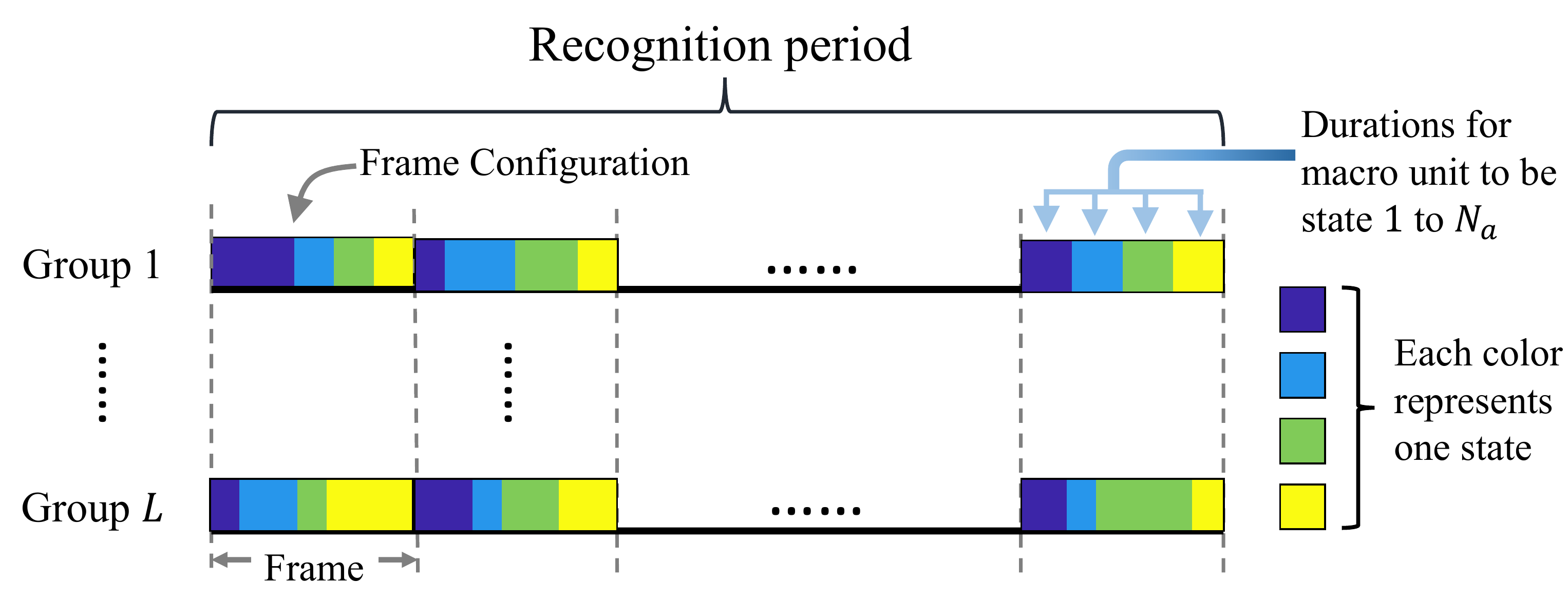}}
	\vspace{-1em}
	\caption{Examples of frame, frame configurations and recognition period with $N_a = 4$.}
	\label{fig: protocol}
\end{figure}

Based on this, we can define the \emph{frame configuration} of the RIS by vector $\bm t=(\tilde{\bm t}_{1}^T, ..., \tilde{\bm t}_{L}^T)^T $, which has size $L\cdot N_a$ and indicates the time duration for $L$ groups to be at the $N_a$ states.
In the RIS-based posture recognition system, the reflected waveforms of the RIS needs to be carefully designed through designing frame configurations, so that the human postures can be recognized with high accuracy.
To alleviate the complexity of the frame configuration design for the RIS, we propose the periodic frame configurations of the RIS.
We define the \emph{recognition period} as the time interval that the sequence of frame configurations of the RIS repeats.
As illustrated in Fig.~\ref{fig: protocol}, the recognition period is composed of $K$ frames.
The $K$ frame configurations in the recognition period is referred to as the \emph{configuration matrix}, which can be expressed as $\bm T = (\bm t_1, ... \bm t_K)^T$, where $\bm t_k = (\tilde{\bm t}_{k,1}^T, ..., \tilde{\bm t}_{k,L}^T)^T$~($k\in[1,K]$) is the configuration of the RIS in the $k$-th frame of the recognition period with $\tilde{\bm t}_{k,l}$~($l\in[1,L]$) being the frame configuration of the $l$-th group in the $k$-th frame.

To recognize the human postures in a recognition period, the receiver measure the $K$ mean values of the signals received in the $K$ frames.
Based on~(\ref{equ: main channel gain}) and~(\ref{equ: received signal}), in the $k$-th frame, the mean value of the received signal can be computed as 
\begin{align}
\label{equ: received signal}
y_k &= h_d\cdot P_t \cdot x + \sum_{m=1}^M\sum_{l=1}^L\sum_{n\in\mathcal N_l} \sum_{i\in\mathcal S_a}t_{k,l,i}\cdot h_{n,m}(\hat{s}_i,\eta_m) \cdot P_t \cdot x + \bar{h}_{\rl}\cdot P_t \cdot x +\bar{\sigma} \nonumber \\
&=h_d\cdot P_t \cdot x + P_t\cdot x\cdot \bm t_k^T\bm A \bm\eta+ \bar{h}_{\rl}\cdot P_t \cdot x +\bar{\sigma}.
\end{align}
Here, $\bm A = (\bm \alpha_1,\dots, \bm \alpha_M)$, where $\bm \alpha_m = (\hat{\bm \alpha}_{m,1}^T, \dots, \hat{\bm \alpha}_{m,L}^T)^T$ with $\hat{\bm \alpha}_{m,l} = (\hat{\alpha}_{m,l,1},\dots,\hat{\alpha}_{m,l,N_a})^T$ and $\hat{\alpha}_{m,l,i} = \sum_{n\in\mathcal N_l}  \lambda\cdot r_{n,m}(\nu_l)\cdot \sqrt{g_{T,n}g_{R,m}}\cdot e^{-j2\pi d_{n,m}/\lambda}/(4\pi d_{n,m}),$ ($l\in[1,L],~i\in[1,N_a]$),
 Besides, $\bar{h}_{rl}$ and $\bar{\sigma}$ denote the average gains of the multi-path component and the noise signal in one frame.

The receiver recognizes the human postures based on the measured $K$ mean values, which constitute a \emph{measurement vector} and can be denoted as $\bm y(\bm T, \bm \eta) = (y_1, ..., y_{K})^T$ or $\bm y$ for simplicity.
Based on~(\ref{equ: received signal}), the measurement vector can be expressed as
\beq
\bm y = h_d\cdot P_t \cdot \bm x + P_t\cdot \bm x\cdot \bm T\bm A \bm\eta+ \bar{h}_{\rl}\cdot P_t \cdot \bm x +\bar{\bm \sigma}.
\eeq
 Since matrix $\bm T\bm A$ determines how the information of human postures is mapped to the measurement, we denote $\bm \Gamma = \bm T\bm A$ and refer to $\bm \Gamma$ as the \emph{measurement matrix}.
More specifically, $\bm y$ is a $K$-dimensional complex vector, i.e., $\bm y\in\mathbb C^{K}$.
Given measurement vector $\bm y$ and configuration matrix $\bm T$, the receiver adopts a likelihood function to perform the posture recognition, which is referred to as the \emph{decision function} and can be expressed as $\mathcal L(\mathcal H_i|\bm y)^T$, $i\in[1,N_P]$.
Here, $\mathcal H_i$~($i\in[1,N_P]$) denotes the hypothesis for the human posture to be the $i$-th posture, and $N_P$ denotes the total number of human postures.
The values of the likelihood function give the probabilities for the receiver to determine the human posture as different postures.

It can be observed that the performance of the posture recognition depends on the design of the decision function as well as the configuration matrix, which need to be optimized in order to obtain high recognition accuracy.
We will formulate the optimizations for the configuration matrix and the decision function in Section~\ref{sec: problem formulation} and propose algorithms to solve the formulated optimization problems in Section~\ref{sec: algorithm design}.

\ifx\allfiles\undefined
\end{document}
\fi

\fi

\ifBreakPage
\newpage
\fi

\ifShowIntro
\ifx\allfiles\undefined
\documentclass[onecolumn,journal,draftclsnofoot,12pt]{IEEEtran}

\usepackage{bm}
\usepackage{amssymb}
\usepackage{amsmath}
\usepackage{amsfonts}
\usepackage{graphicx}
\usepackage{subfigure}
\usepackage{cite}
\usepackage{enumerate}
\usepackage{url}
\usepackage{epstopdf}
\usepackage{float}
\usepackage{color}
\usepackage{multirow}

\usepackage[normalem]{ulem}  
\newcommand{\beq}{\begin{equation}}
\newcommand{\eeq}{\end{equation}}
\newcommand{\blue}{\textcolor[rgb]{0,0,0.7}}
\newcommand{\pasted}{\textcolor[rgb]{0.3,0,0}}
\allowdisplaybreaks[4]
\makeatletter  
\newif\if@restonecol  
\makeatother  
\let\algorithm\relax  
\let\endalgorithm\relax  
\usepackage[linesnumbered, lined, commentsnumbered, ruled]{algorithm2e}
\usepackage{amsmath}  

\usepackage{makecell} 

\renewcommand{\Pr}{\mathrm{Pr}}
\newcommand{\diag}{\mathrm{diag}}
\newcommand{\rank}{\mathrm{rank}}
\newcommand{\tx}{\mathrm{Tx}}
\newcommand{\rx}{\mathrm{Rx}}
\newcommand{\rl}{\mathrm{rl}}
\newcommand{\mau}{\mathrm{mu}}
\newcommand{\sign}{\mathrm{sign}}

\DeclareMathOperator*{\argmin}{argmin}
\DeclareMathOperator*{\argmax}{argmax}

\newif\ifshowannote
\showannotetrue

\newtheorem{definition}{\bf Definition}
\newtheorem{proposition}{\bf Proposition}
\newtheorem{theorem}{\bf Theorem}
\newtheorem{lemma}{\bf Lemma}
\newtheorem{proof}{Proof}

\begin{document}
\fi
\section{Problem Formulation of RIS-based Posture Recognition}
\label{sec: problem formulation}
In this section, we formulate the problem to optimize the RIS-based posture recognition system by minimizing the average cost of false posture recognition.
We then decompose it into two subproblems, i.e., the configuration optimization and the decision function optimization.

\subsection{Problem Formulation}
For the $i$-th posture, let us denote by $\bm \eta_i$ the corresponding space reflection vector.
The measurement vector for the $i$-th posture can be expressed as $\bm y_i = \bm y ( \bm T, \bm \eta_i)$.
Since the objective of the system is to recognize human postures accurately, we minimize the average cost due to the false recognition of the system.
The objective function, which we refer to as the \emph{average false recognition cost}, can be calculated as
\beq
\label{equ: classification cost2}
\Psi_{\mathcal L} =\sum_{i, i'\in\mathcal [1,N_p],~i\neq i'} p_i\cdot  \chi_{i,i'}\cdot \int_{\bm y_i\in\mathbb C^{K}} \Pr(\bm y_i|\bm \eta_i) \cdot \mathcal L\left( \mathcal H_{i'}| \bm y_i \right)\cdot  d\bm y_i,
\eeq
where
	$p_i\in[0,1]$ denotes the probability for the $i$-th human posture to appear and $\sum_{i\in[1,N_P]} p_i = 1$,
	$\chi_{i,i'}\in \mathbb R^+$ denotes the \emph{cost} for the false recognition of actual $i$-th posture as the $i'$-th posture,
	and $\Pr(\bm y_i|\bm \eta_i)$ denotes the probability for the measurement vector to be $\bm y_i$, given space reflection vector $\bm\eta_i$.

Based on~(\ref{equ: received signal}) and~(\ref{equ: classification cost2}), the optimization for the RIS-based posture recognition system can be formulated as
\begin{align}
\label{opt: falling recognizing obj}
(\text{P1}) : 
\min_{
\bm T, \mathcal L
 }
~
&\Psi_{\mathcal L},\\
\label{opt: falling recognizing st1}
 s.t. ~
 &  \mathcal L\left( \mathcal H_{i'}| \bm y_i\right) \geq 0,~
	\forall \bm y_i \in\mathbb C^{K},~i,i'\in[1,N_a],\\
\label{opt: falling recognizing st2}
 & \sum_{i\in[1,N_P]}\mathcal L\left(\mathcal H_{i'}|\bm y_i \right) = 1, ~
	\forall \bm y\in\mathbb C^{K},\\
\label{opt: falling recognizing st3}
&\bm y_i =h_d\cdot P_t \cdot \bm x + P_t\cdot \bm x\cdot \bm T\bm A \bm\eta_i+ \bar{h}_{\rl}\cdot P_t \cdot \bm x +\bar{\bm \sigma}, ~\forall i \in[1,N_P],\\
\label{opt: falling recognizing st4}
&\bm 1^T\tilde{ \bm t}_{k,l} = \delta,~\forall k\in[1,K], l\in[1,L],\\
\label{opt: falling recognizing st5}
&\tilde{t}_{k,l,i}\geq 0,~\forall k\in[1,K], l\in[1,L], i\in[1,N_a].
\end{align}
In~(P1), the optimization variables are the decision function, i.e., $\mathcal L$, and the configuration matrix, i.e., $\bm T$.
Constraints~(\ref{opt: falling recognizing st1}) and~(\ref{opt: falling recognizing st2}) are due to the fact that the decision function returns probabilities.
Constraint~(\ref{opt: falling recognizing st3}) denotes the relationship between measurement vector $\bm y_i$ and space reflection vector of the $i$-th posture $\bm \eta_i$.
Constraints~(\ref{opt: falling recognizing st4}) and~(\ref{opt: falling recognizing st5}) indicate that the time duration for all the states is positive and sum up to $\delta$ in each frame.

\subsection{Problem Decomposition}
In~(P1), the configuration matrix and the decision function are coupled and need to be optimized jointly, which makes~(P1) hard to solve.
To handle this difficulty, we decompose~(P1) into two sub-problems by separating the configuration matrix optimization and the decision function optimization.
The two sub-problems are referred to as the \emph{configuration matrix optimization problem} and the \emph{decision function optimization problem}, which are described as follows.
\subsubsection{Configuration Matrix Optimization}
Given a decision function, the optimization problem for configuration matrix $\bm T$ can be formulated as
\begin{align}
\label{opt: configuration optimization}
(\text{P2}) : 
\min_{
 \bm D
 }
\quad 
&\Psi_{\mathcal L},\\
\label{opt: config design opt st1}
 s.t. \quad
&\bm y_i =h_d\cdot P_t \cdot \bm x + P_t\cdot \bm x\cdot \bm T\bm A \bm\eta_i+ \bar{h}_{\rl}\cdot P_t \cdot \bm x +\bar{\bm \sigma}, ~\forall i \in[1,N_P],\\
\label{P2st: 3}
&\bm 1^T\tilde{ \bm t}_{k,l} = \delta,~\forall k\in[1,K], l\in[1,L],\\
\label{P2st: 4}
&\tilde{t}_{k,l,i}\geq 0,~\forall k\in[1,K], l\in[1,L], i\in[1,N_a].
\end{align}


\subsubsection{Decision Function Optimization}
Given a configuration sequence, the optimization for the decision function $\mathcal L$ can then be formulated as
\begin{align}
\label{opt: decision function optimization}
(\text{P3}) : 
\min_{
\mathcal L
}
\quad 
&\Psi_{\mathcal L}, \\
\label{opt: decision function optimization st1}
 s.t. \quad
 &  \mathcal L\left( \mathcal H_{i'}| \bm y_i\right) \geq 0,~
	\forall \bm y_i \in\mathbb C^{K},~i,i'\in[1,N_a],\\
\label{opt: decision function optimization st2}
 &\sum_{i\in[1,N_P]}\mathcal L\left(\mathcal H_{i'}|\bm y_i \right) = 1, ~
	\forall \bm y\in\mathbb C^{K}.
\end{align}

In the following Section~\ref{sec: algorithm design}, we design the algorithms to solve the configuration matrix optimization and the decision function optimization problems, respectively.
\ifx\allfiles\undefined
\end{document}
\fi
\fi
\ifBreakPage
\newpage
\fi

\ifBreakPage
\newpage
\fi

\ifShowIntro
\ifx\allfiles\undefined
\documentclass[onecolumn,journal,draftclsnofoot,12pt]{IEEEtran}

\usepackage{bm}
\usepackage{amssymb}
\usepackage{amsmath}
\usepackage{amsfonts}
\usepackage{graphicx}
\usepackage{subfigure}
\usepackage{cite}
\usepackage{enumerate}
\usepackage{url}
\usepackage{epstopdf}
\usepackage{float}
\usepackage{color}
\usepackage{multirow}

\usepackage[normalem]{ulem}  
\newcommand{\beq}{\begin{equation}}
\newcommand{\eeq}{\end{equation}}
\newcommand{\blue}{\textcolor[rgb]{0,0,0.7}}
\newcommand{\pasted}{\textcolor[rgb]{0.3,0,0}}
\allowdisplaybreaks[4]
\makeatletter  
\newif\if@restonecol  
\makeatother  
\let\algorithm\relax  
\let\endalgorithm\relax  
\usepackage[linesnumbered, lined, commentsnumbered, ruled]{algorithm2e}
\usepackage{amsmath}  

\usepackage{makecell} 

\renewcommand{\Pr}{\mathrm{Pr}}
\newcommand{\diag}{\mathrm{diag}}
\newcommand{\rank}{\mathrm{rank}}
\newcommand{\tx}{\mathrm{Tx}}
\newcommand{\rx}{\mathrm{Rx}}
\newcommand{\rl}{\mathrm{rl}}
\newcommand{\mau}{\mathrm{mu}}
\newcommand{\sign}{\mathrm{sign}}

\DeclareMathOperator*{\argmin}{argmin}
\DeclareMathOperator*{\argmax}{argmax}

\newif\ifshowannote
\showannotetrue

\newtheorem{definition}{\bf Definition}
\newtheorem{proposition}{\bf Proposition}
\newtheorem{theorem}{\bf Theorem}
\newtheorem{lemma}{\bf Lemma}
\newtheorem{proof}{Proof}

\begin{document}
\fi
\section{Algorithms for Configuration Matrix and Decision Function Optimizations}
\label{sec: algorithm design}
In this section, we first propose the algorithms to solve the configuration optimization and decision function optimization.
Then, we analyze the convergence of the proposed algorithms.

\subsection{Configuration Optimization Algorithm}
\label{sec: config. design opt.}
In~(P2), we optimize the configuration matrix to minimize the average false recognition cost.
Solving~(P2) explicitly requires the space reflection vectors, i.e., $\bm \eta_i$~($i\in[1.N_P]$) to be known in prior.
Besides, the optimized configuration matrix for specific known coefficient vectors may be sensitive to the subtle changes of the postures.
Therefore, instead of optimizing RIS configurations given specific space reflection vectors, we will find an optimal configuration matrix for the general posture recognition scenarios.

As the decision function recognizes the human postures based on measurement vector $\bm y$, we consider optimizing the configuration matrix $\bm T$ so that $\bm y$ is able to carry the richest information about the human postures.
As indicated in (\ref{opt: config design opt st1}), the information of human body distribution is contained in space reflection vector $\bm\eta$.
Therefore, intuitively, it requires that $\bm \eta$ can be potentially reconstructed from $\bm y$ with the minimum loss.

Since the signals from the multi-path component and the noise are usually much smaller than the reflection channels gains and are random values determined by the environment, we neglect them and consider the relation between $\bm \eta$ and $\bm y$ as $\bm y =P_t\cdot \bm x\cdot \bm T \bm A\bm \eta$.
Since the number of space blocks is large, we assume $M\gg K$.
In this case, $\bm y =P_t\cdot \bm x\cdot \bm T\bm A \bm \eta$ is an underdetermined equation which has an infinite number of solutions.
Therefore, the true $\bm \eta$ cannot be reconstructed from $\bm y$ given $\bm T\bm A$, unless additional constraints on $\bm \eta$ are employed.

One of the usually employed constraints to reconstruct the target signals in an underdetermined equation is that the signal to reconstruct is \emph{sparse}.
Intuitively, the target signal, i.e., $\bm \eta$, is sparse when the number of its nonzero entries is sufficiently smaller than the dimension of the signal vector, i.e., 
\beq
\label{equ: sparse condition}
|\mathrm{supp}(\bm \eta)| \ll \mathrm{dim}(\bm \eta).
\eeq
Here, $\mathrm{dim}(\bm \eta)$ denotes the dimension of $\bm \eta$,
	$\mathrm{supp}(\bm \eta) = \{\eta_i| \eta_i \neq 0, i\in[1,\mathrm{dim }(\bm\eta) ]\}$ indicates the support set of $\bm \eta$,
	and $|\cdot|$ provides the cardinality of a set~\cite{eldar2012compressed}. 
From Fig.~\ref{fig: channel_model}, it can be observed that in the proposed fall-detection system, most of the space blocks are empty and thus have zero reflection coefficients.
Besides, for the space blocks where the human body lies, only those that contain the surfaces of the human body with specific angles can reflect the incidence signals towards the receiver and have non-zero reflection coefficients.
Therefore, $\bm \eta$ is a space vector, and condition~(\ref{equ: sparse condition}) is satisfied.
The sparse target signals in an underdetermined equation can be reconstructed efficiently using the approaches such as \emph{compressive sensing}~\cite{han_COMPRESSIVE}.

Based on~\cite{elad2007optimized}, to minimize the loss of reconstruction for sparse target signals, we can minimize the averaged \emph{mutual coherence} of measurement matrix $\bm \Gamma = \bm T\bm A$, which is defined as
\beq
\label{equ: mutual coherence}
\mu(\bm \Gamma)= \frac{1}{M(M-1)}\cdot\sum_{m,m'\in[1,M],m\neq m'}\frac{|\bm \gamma_m^T \bm\gamma_{m'}|}{\|\bm \gamma_m\|_2\cdot \|\bm \gamma_{m'}\|_2},
\eeq
where $\bm \gamma_m = \bm T\bm\alpha_m$ denotes the $m$-th column of $\bm \Gamma$, and $\|\cdot\|_{2}$ denotes the $l_{2}$-norm.

Therefore, based on~(\ref{equ: mutual coherence}), we can reformulate the configuration sequence optimization as the following mutual coherence minimization problem.
\begin{align}
(\text{P}4)~
\min_{\bm T}~
& \mu(\bm \Gamma),  \\
s.t. \quad 
\label{opt: reform 6 st1}
&\bm \gamma_m = \bm T\bm \alpha_m, ~\forall m\in[1,M],\\
\label{opt: reform 6 st2}
&\bm 1^T\tilde{ \bm t}_{k,l} = \delta,~\forall k\in[1,K], l\in[1,L],\\
\label{opt: reform 6 st3}
&\tilde{t}_{k,l,i}\geq 0,~\forall k\in[1,K], l\in[1,L], i\in[1,N_a].
 \end{align}

Due to the non-convex objective function, (P6) is a non-convex optimization problem and NP-hard.
Besides, it can be observed that the number of variables of~(P4) is $K\cdot L\cdot N_a $.
In practical scenarios, the number of measurements and the number of groups can be large, which makes $K\cdot L \cdot N_a$ a large number.
Moreover, the variables of~(P4) are coupled together, and thus~(P4) cannot be separated into independent sub-problems, resulting in prohibitive computational complexity.

To solve~(P4) in an acceptable complexity in practice, we propose a low-complexity algorithm to solve (P4) sub-optimally based on the alternating optimization~(AO) technique.
The proposed algorithm is referred to as the frame configuration alternating optimization~(FCAO) algorithm.
In the FCAO algorithm, we alternately optimize each of the frame configuration in an iterative manner by fixing the other $K-1$ frame configurations, until the convergence is achieved.

In each iteration, we need to optimize~(P4) with respect to $\bm t_k$, with $\bm T_{-k}\!=\! (\bm t_1,...,\bm t_{k-1},\bm t_{k+1},...,\bm t_K)$ fixed.
Besides, we denote the coherence of $\bm \gamma_{m}$ and $\bm \gamma_{m'}$ by $u_{m,m'}$, i.e.,
\beq
u_{m,m'} = \frac{\bm \gamma_m^T \bm\gamma_{m'}}{\|\bm \gamma_m\|_2\cdot \|\bm \gamma_{m'}\|_2},
\eeq
and arrange $u_{m,m'}$~($m,m'\in [1,M],m\neq m'$) as vector $\bm u = (u_{1,2},...,u_{1,M},...,u_{2,3},...,u_{M-1,M})$.
The optimization problem for each iteration can then be formulated as
\begin{align}
(\text{P}5)~
\min_{\bm t_k, \bm u}~
& \|\bm u\|_1,\\
s.t. \quad 
\label{opt: reform 5 st1}
& u_{m,m'} = \frac{\bm \gamma_m^T \bm\gamma_{m'}}{\|\bm \gamma_m\|_2\cdot \|\bm \gamma_{m'}\|_2},~m,m'\in[1,M],m\neq m',  \\
\label{opt: reform 5 st2}
&\bm 1^T \tilde{\bm t}_{k,l} = \delta, ~\forall l\in[1,L],\\
\label{opt: reform 5 st3}
&\tilde{\bm t}_{k,l}\succeq 0,~\forall l\in[1,L],
 \end{align}
 where $\|\cdot\|_1$ denotes the $l_1$-norm of the contained vector.

To solve (P5), we can adopt the augmented Lagrangian  method~\cite{Migliore2011Compressed}, where the original constrained optimization problem is handled by solving a sequence of unconstrained minimizations for the augmented Lagrangian function.
To express the augmented Lagrangian function, we first define the indicator function for (P5) as 
\beq
\mathbb I(\bm t_k) = \begin{cases}
 	&1, \quad \text{if }\tilde{\bm t}_{k,l}\succeq 0 \text{ and } \bm 1^T\tilde{\bm t}_{k,l} = 1,~\forall l\in[1,L],\\
 	&0, \quad \text{otherwise}.
 \end{cases}
\eeq
The augmented Lagrangian function for~(P6) can be expressed as 
\begin{align}
\label{equ: convenient aug. Lagrangian}
\mathcal L_{A}&(\bm t_k, \bm u; \bm \beta, \rho, \bm T_{-k})  =  \|\bm u\|_{1} + \mathbb I(\bm t_k) \\
& + \sum_{\substack{m,m'\in[1,M], \\ m\neq m'}} \beta_{m,m'} \left(u_{m,m'} - \frac{\bm \gamma_m^T \bm\gamma_{m'}}{\|\bm \gamma_m\|_2\cdot \|\bm \gamma_{m'}\|_2}\right) 
+ \sum_{\substack{m,m'\in[1,M], \\ m\neq m'}} \frac{\rho}{2} \left| u_{m,m'} -\frac{\bm \gamma_m^T \bm\gamma_{m'}}{\|\bm \gamma_m\|_2\cdot \|\bm \gamma_{m'}\|_2}\right|^2, \nonumber
\end{align}
where $\rho$ is a positive scaling factor and $\bm \beta = (\beta_{1,2}, \beta_{1,3}, \dots, \beta_{1,M}, \beta_{2,3},\dots, \beta_{M-1,M}) \in \mathbb C^{M(M-1)/2}$ is the vector for Lagrange multipliers.
The augmented Lagrangian method finds a local optimal solution to~(P5) by minimizing a sequence of~(\ref{equ: convenient aug. Lagrangian}) where $\bm \beta$ is fixed in each iteration.
Specifically, the sequence of unconstrained Lagrangian minimization can be solved using an \emph{alternating minimization procedure},  in which $\bm u$ is updated while $\bm t_k$ is fixed and vice versa.
The completed algorithm to solve~(P6) by the augmented Lagrangian method can be found in Appendix~\ref{appx: augmented Lagrangian alg.}.

However, the augmented Lagrangian method may result in a local optimum far from the global optimum, if the starting point is badly chosen.
Therefore, we adopt an intuitive algorithm named \emph{pattern search}~\cite{lewis2007implementing}, which can obtain a good initial point for the augmented Lagrangian method.
The complete FCAO algorithm for~(P4) can be summarized as Algorithm~\ref{alg: ao algorithm}.

\begin{algorithm}[!t]  \label{alg: ao algorithm}
\small
\caption{FCAO algorithm for solving (P4)}
	\SetKwInOut{Input}{Input}
	\SetKwInOut{Output}{Output}
\Input{Initial random feasible configuration matrix $\bm T^{(0)}$.}
\Output{
Optimal averaged mutual coherence $\mu^*$ and configuration matrix $\bm T^*$ for (P4).}

Compute initial $ \mu^{(0)}$ based on~(\ref{opt: reform 6 st1}) given $\bm T^{(0)}$\;

Set the number of consecutive iterations with no improvements as $N_{\mathrm{non}} = 0$ and current frame index $k=1$\;

\For{$i = 1,2,...$}
{
	Invoke pattern search method in~\cite{lewis2007implementing} for~(P5) to obtain a initial $\bm t_k'$ which results in low average mutual coherence given fixed $\bm T^{(i-1)}_{-k}$\;
	
	Using $\bm t_k$ as an initial point, solve (P5) by using the augmented Lagrangian method in described in Appendix~\ref{appx: augmented Lagrangian alg.}, and denote the resulting minimum mutual coherence as $\mu'$ and optimal configuration as $\bm t_{k}'$\;
	
	If $\mu'<\mu^{(i)} $, update $\mu^{(i)} = \mu'$, the $k$-th frame configuration in $\bm T^{(i)}$ as $\bm t^{(i)}_k = \bm t_k'$, $\bm T^{(i)}_{-k} = \bm T^{(i-1)}_{-k}$, and $N_{\mathrm{non}} = 0$; otherwise, set $N_{\mathrm{non}} = N_{\mathrm{non}} + 1$\;
	
	If $N_{\mathrm{non}} < K$, set $k = \mathrm{mod}(k+1, K) + 1$; otherwise, return $\mu^* = \mu^{(i)}$ and $\bm T^* = \bm T^{(i)}$\;
}

\end{algorithm}
%
\subsection{Supervised learning algorithm for solving (P6)}
\label{ssec: decision func. opt.}

\begin{algorithm}[!t]  \label{alg: back-propagation algorithm}
\small
\caption{Supervised learning algorithm for solving~(P6)}
	\SetKwInOut{Input}{Input}
	\SetKwInOut{Output}{Output}
\Input{
Training data set $\mathcal D = \{(\bm y_j, L_j)\}$.
Learning rate $\zeta$.
}
\Output{
Trained parameter $\bm \theta^{*}$, which is the solution for (P6).
}
Obtain initial $\bm \theta^{(0)}$ with random value which follows uniform distribution within $(0,1)$\;
Calculate the current average recognition cost for the training data set $\Psi^{(0)}= \sum_{j=1}^{N_d}E_j$ based on~(\ref{equ: loss function})\;
\For{i = 1,2,...}
{
\For{$(\bm y_j, L_j) \in \mathcal D$}
{
Input $\bm y_j$ to the NN $\mathcal L^{\bm \theta}|_{\bm\theta = \bm\theta^{(i-1)}}$ and obtain output $\tilde{p}_j$ \;
Calculate the gradient of the loss with respect to the current parameter $\bm \theta$, i.e. $\partial E_j/ \partial \bm \theta$\;
Update parameter by $\bm \theta'= \bm \theta^{(i-1)} + \Delta\bm\theta$ based on~(\ref{equ: adjustment})\;
}
Calculate $\Psi'= \sum_{j=1}^{N_d}E_j$ based on~(\ref{equ: loss function}) using the updated $\mathcal L^{\bm \theta}$\;
If $\Psi' \geq \Psi^{(i-1)}$, break and output $\bm\theta^* = \bm \theta^{(i-1)}$; otherwise, set $\Psi^{(i)} =  \Psi'$ and $\bm \theta^{(i)} = \bm \theta'$.
}
\end{algorithm}

To solve~(P3) efficiently, we parameterize $\mathcal L$ by a real-valued parameter vector $\bm \theta\in\mathbb R^L$.
Let us denote the parameterized function as $\mathcal L^{\bm \theta}$, and the decision function optimization problem can be formulated as
\begin{align}
\label{opt: decision function optimization}
(\text{P6}) : 
\min_{
\bm \theta
}
\quad 
&\Psi_{\bm \theta}(\{\bm s_k\}_{k=1}^K), \\
\label{opt: decision function optimization st1}
 s.t. \quad
 &  \mathcal L^{\bm \theta}\left( \mathcal H_i| \bm y\right) \geq 0,~
	\forall \bm y\in\mathbb C^{K},~i\in[1,N_p],\\
\label{opt: decision function optimization st2}
 & \sum_{i\in[1,N_a]}\mathcal L^{\bm\theta}\left(\mathcal H_i|\bm y \right) = 1, ~
	\forall \bm y\in\mathbb C^{K}.
\end{align}

To solve~(P6), we propose the following algorithm based on the supervised learning approach using neural network~(NN)~\cite{goodfellow2016deep}.
In the designed human posture recognition system, we adopt a fully connected NN for the decision function $\mathcal L^{\bm \theta}$.
The fully connected NN consists of the input layer, hidden layers, and the output layer, which are connected successively.
The input layer takes the $K$-dimensional complex-valued measurement vector $\bm y$ in a recognition period and passes it to the first hidden layer.
The $j$-th hidden layer has $n_{\text{hid}, j}$ nodes.
Each node calculates a biased weighted sum of its input, processes the sum with an \emph{activation function}, and outputs the result to the nodes connected to it in the next layer.
In the output layer, the number of nodes is number of posture $N_p$. 
The output nodes handle their input with a \emph{softmax} function $f_{\text{softmax}}$~\cite{bishop2006pattern}, which can be expressed as 
\beq
\bm y  = f_{\text{softmax}}(\bm x) = \begin{bmatrix}
\frac{e^{x_1}}{\sum_{n=1}^{N_p} e^{x_n}}, &\dots, &  \frac{e^{x_{N_p}}}{\sum_{n=1}^{N_p} e^{x_n}}
 \end{bmatrix}.
\eeq
It can be observed that the softmax function converts the input of the output layer $\bm x$ to $\bm y$ which satisfies constraints~(\ref{opt: decision function optimization st1}) and~(\ref{opt: decision function optimization st2}) and can be considered as the probabilities for the $N_p$ postures.
In the NN, the parameters for the decision function, i.e., $\bm \theta$, stands for the weights of the connections and the biases of the nodes.

It can be seen that parameter $\bm \theta$ in NN determines how the input is processed into output and thus determines the performance of the decision function in terms of average false recognition cost~(\ref{opt: decision function optimization}).
Therefore,~(P5) is equivalent to finding the optimal $\bm \theta$ for the NN, which minimize~(\ref{opt: decision function optimization st1}).
To solve~(P5), we propose the following algorithm based on the back-propagation algorithm~\cite{pineda1987generalization}.
To train the NN requires a training data set of the measurement vectors labeled with the postures, which can be expressed as 
\beq
\mathcal D = \{(\bm y_j, L_j)\}_{j=1}^{N_{data}}.
\eeq
Here, $\bm y_j$ and $L_j$ denote the $j$-th measurement vector and its label, i.e., the index of posture, respectively, and $N_{data}$ denotes the size of data set.
The data set is collected in prior for a given configuration sequence.
Besides, the number of labeled data for the $i$-th posture needs to be approximately $p_i\cdot N_{data}$, where $p_i$ is defined in~(\ref{equ: classification cost2}).

For $\bm y_j$ in a data pair $(\bm y_j, L_j)$, we assume that the output of the NN is denoted by $\tilde{p}_j$, which is a $N_p$-dimensional vector.
The $i$-th element in $\tilde{p}_j$ is the probability that the posture is detected as the $i$-th posture.
In accordance with~(\ref{equ: classification cost2}), for $(\bm y_j, L_j)$, the loss function, i.e., the recognition cost incurred by the NN is defined as
\beq
\label{equ: loss function}
E_j = \sum_{i=1}^{N_p}\sum_{i' = 1}^{N_p} \tilde{p}_{j,i'} \chi_{i,i'}I_{j,i},
\eeq
where $\bm I_j = (I_{j,1},...,I_{j,N_p})$ is a one-hot indicator vector for the $j$-th data.
To be specific, $I_{j,i'} = 1$ if $i' = L_j$; otherwise, $I_{j,i'} = 0$.
Based on the back-propagation algorithm, the parameters of the NN need to be updated in the negative gradient direction of the loss function~(\ref{equ: loss function}). 
The adjustment of $\bm \theta$ can be expressed as
\beq
\label{equ: adjustment}
\Delta \bm \theta = -\zeta \frac{\partial E_j}{\partial \bm \theta},
\eeq
where $\eta\in(0,1)$ denotes the learning rate.
The update of parameter $\bm \theta$ proceeds iteratively and repeatedly until the average loss, i.e., the average recognition cost due to the NN over the data set, converges.
The algorithm to solve~(P6) is summarized as Algorithm~\ref{alg: back-propagation algorithm}.


\subsection{Convergence Analysis}
In the following, we analyze the convergence of the proposed algorithms in this section.
\subsubsection{Convergence of FCAO Algorithm}
Based on Algorithm~\ref{alg: ao algorithm}, in the $(i+1)$-th iteration, a better configuration matrix which results in lower $\mu$ can be obtained given configuration matrix $\bm T^{(i)}$ in the $(i+1)$-th iteration.
Therefore, we have $
\mu(\bm T^{(i+1)}\bm A) \leq \mu(\bm T^{(i)}\bm A),
$
which implies that the objective value obtained in Algorithm~\ref{alg: ao algorithm} is non-increasing after each iteration of the FCAO algorithm.
Since the average mutual coherence of $\bm\Gamma = \bm T\bm A$ has lower-bound of $0$, the proposed FCAO algorithm is guaranteed to converge.

\subsubsection{Convergence of Supervised Learning Algorithm}
Similar to the convergence analysis of the FCAO algorithm, in the $(i+1)$-th iteration, parameter $\bm \theta$ only updates to $\bm\theta'$ when the resulting average cost $\psi'<\psi^{(i)}$. 
Therefore, the objective value obtained in Algorithm~\ref{alg: back-propagation algorithm} is also non-increasing.
As the average false recognition cost is lower bounded with zero, the proposed Supervised learning is guaranteed to converge.

\ifx\allfiles\undefined
\end{document}
\fi
\fi
\ifBreakPage
\newpage
\fi

\ifBreakPage
\newpage
\fi

\ifShowSimul
\ifx\allfiles\undefined
\documentclass[onecolumn,journal,draftclsnofoot,12pt]{IEEEtran}

\usepackage{bm}
\usepackage{amssymb}
\usepackage{amsmath}
\usepackage{amsfonts}
\usepackage{graphicx}
\usepackage{subfigure}
\usepackage{cite}
\usepackage{enumerate}
\usepackage{url}
\usepackage{epstopdf}
\usepackage{float}
\usepackage{color}
\usepackage{multirow}

\usepackage[normalem]{ulem}  
\newcommand{\beq}{\begin{equation}}
\newcommand{\eeq}{\end{equation}}
\newcommand{\blue}{\textcolor[rgb]{0,0,0.7}}
\newcommand{\pasted}{\textcolor[rgb]{0.3,0,0}}
\allowdisplaybreaks[4]
\makeatletter  
\newif\if@restonecol  
\makeatother  
\let\algorithm\relax  
\let\endalgorithm\relax  
\usepackage[linesnumbered, lined, commentsnumbered, ruled]{algorithm2e}
\usepackage{amsmath}  

\usepackage{makecell} 

\renewcommand{\Pr}{\mathrm{Pr}}
\newcommand{\diag}{\mathrm{diag}}
\newcommand{\rank}{\mathrm{rank}}
\newcommand{\tx}{\mathrm{Tx}}
\newcommand{\rx}{\mathrm{Rx}}
\newcommand{\rl}{\mathrm{rl}}
\newcommand{\mau}{\mathrm{mu}}
\newcommand{\sign}{\mathrm{sign}}

\DeclareMathOperator*{\argmin}{argmin}
\DeclareMathOperator*{\argmax}{argmax}

\newif\ifshowannote
\showannotetrue

\newtheorem{definition}{\bf Definition}
\newtheorem{proposition}{\bf Proposition}
\newtheorem{theorem}{\bf Theorem}
\newtheorem{lemma}{\bf Lemma}
\newtheorem{proof}{Proof}

\begin{document}
\fi
\section{System Implementation}
\label{sec: system implementation}

In this section, we elaborate on the implementation of the RIS-based posture recognition system.
We first elaborate on the implementation of the RIS, and then describe the implementation of the transceiver module.
	
\subsection{Implementation of RIS}
\label{sec: ris implementation}
\begin{table*}[!t]
\caption{Normal direction forward transmission gain of RIS element in different states}
\vspace{-1.2em}
\centering
\scriptsize
\begin{tabular}{| c | c| c | c | p{4cm}<{\centering}| p{4cm}<{\centering} |}
\Xhline{1.pt}
\multirow{2}*{\textbf{State}} &
\multicolumn{3}{c|}{\textbf{Bias Voltages on PIN Diodes}}  &
\multicolumn{2}{c|}{\textbf{Normal direction forward Transmission Gains in CST Simulation}}\\ 
\cline{2-6} 
	& \textbf{PIN \# 1} &\textbf{PIN \#2} &\textbf{PIN \#3} & \textbf{Phase Shift} & \textbf{Amplitude Ratio}\\
\hline
$\hat{s}_1$ & $0$V~(OFF) & $0$V~(OFF) & $0$V~(OFF) & $\pi/4$ & $0.97$ \\
$\hat{s}_2$ & $0$V~(ON) & $0$V~(OFF) & $3.3$V~(ON) & $3\pi/4$ & $0.97$ \\
$\hat{s}_3$ & $3.3$V~(ON) & $0$V~(OFF) & $3.3$V~(ON)  & $5\pi/4$ & $0.92$ \\
$\hat{s}_4$ & $3.3$V~(ON) & $3.3$V~(ON) & $0$V~(OFF) & $7\pi/4$ & $0.88$ \\
\Xhline{1.pt}
\end{tabular}
\label{table: S21 at 4 states}
\end{table*}

\begin{figure}[!t] 
	\center{\includegraphics[width=0.75\linewidth]{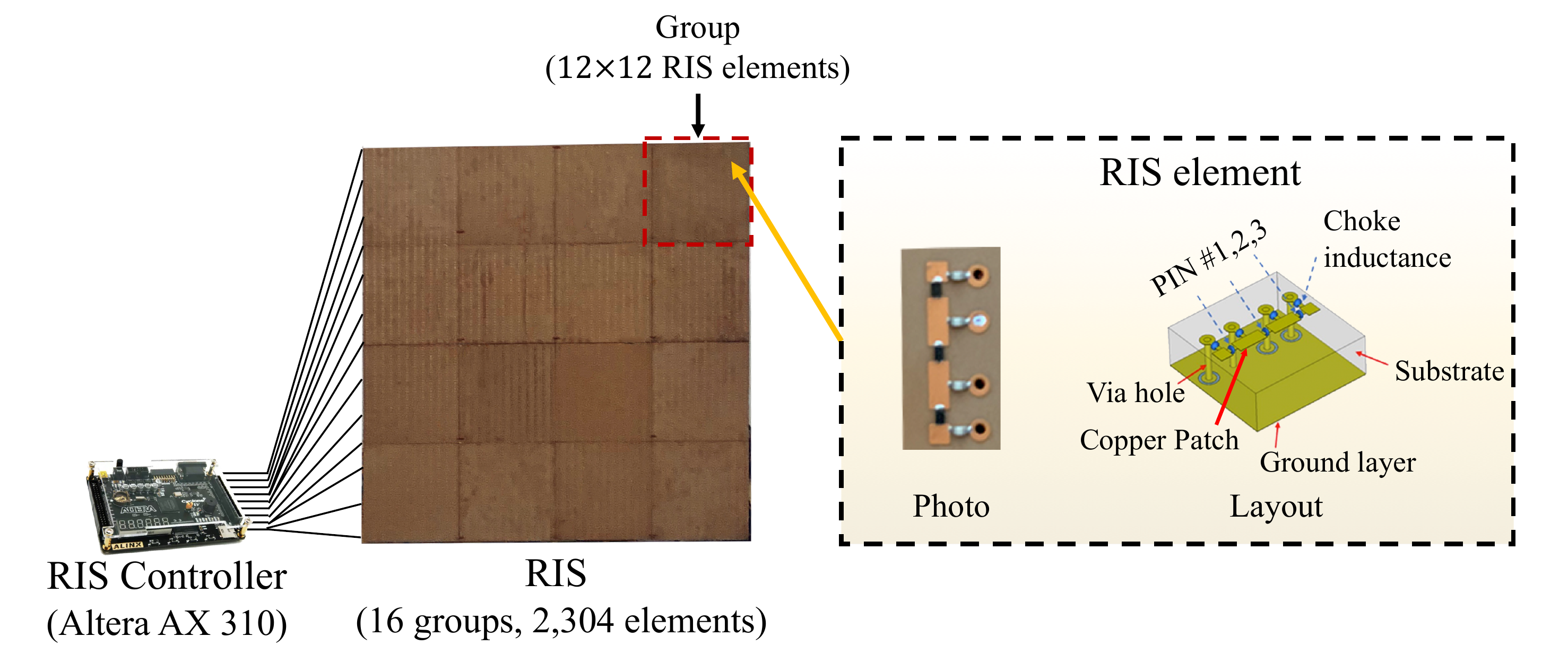}}
	\vspace{-1em}
	\caption{RIS controller, RIS, and RIS element.}
	\label{fig: RIS}
\end{figure}

We adopt the electrically modulated RIS proposed in~\cite{Li2019Machine}, which is shown in Fig.~\ref{fig: RIS}.
The RIS is with the size of $69\times 69\times 0.52$ cm$^3$ and is composed of a two-dimensional array of electrically controllable RIS elements.
Each row/column of the array contains $48$ RIS elements, and therefore, the total number of RIS elements is $2,304$. 

Each RIS element has the size of $1.5\times 1.5 \times 0.52$ cm$^3$ and is composed of $4$ rectangle copper patches printed on a dielectric substrate~(Rogers 3010) with dielectric constant of $10.2$.
Any two adjacent copper patches are connected by a PIN diode~(BAR 65-02L), and each PIN diode has two operation states, i.e., ON and OFF, which are controlled by applied bias voltages on the via holes.
Specifically, when the applied bias voltage is $3.3$V~(or $0$V), the PIN diode is at the ON~(or OFF) state.
Besides, to isolate the DC feeding port and microwave signal, four choke inductors of $30$nH are used in each RIS element. 
Besides, as shown in Fig.~\ref{fig: RIS}, an RIS element contains four choke inductors that are used to isolate the DC feeding port and RF signals.

As there are $N_D = 3$ PIN diodes for an RIS element, the total number of possible states for an RIS element is $8$.
We simulate the $S_{21}$ parameters, i.e., the forward transmission gain, of the RIS element in $4$ selected states for normal-direction incidence RF signals in CST software, Microwave Studio, Transient Simulation Package~\cite{Hirtenfelder2007Effective}. 
Table~\ref{table: S21 at 4 states} provides the amplitude ratios and phase shifts of the RIS element in four selected states for incidence sinusoidal signals with frequency $3.198$~GHz.
It can be observed that the four selected states have the phase shift values equaling to $\pi/4$, $3\pi/4$, $5\pi/4$ and $7\pi/4$, respectively.
We pick these four states with a phase shift interval equaling to $\pi/2$ as the available state set $\mathcal S_a$, i.e., $\mathcal S_a = \{\hat{s}_1,\hat{s}_2,\hat{s}_3,\hat{s}_4\}$.

The RIS elements are divided into $16$ groups, and thus each group contains $12\times 12$ adjacent RIS elements arranged squarely.
The RIS elements within the same group are in the same state.
As shown in Fig.~\ref{fig: RIS}, the states of the $16$ groups are controlled by a \emph{RIS controller}, which is implemented by a field-programmable gate array~(FPGA)~(ALTERA AX301).
Specifically, we use the expansion ports on the FPGA to control the frame configuration of the RIS.
Every three expansion ports control the state of one group by applying bias voltages on the PIN diodes.
Besides, the FPGA is pre-loaded with the configuration sequence matrix in Section~\ref{sec: config. design opt.}.
The configurations of the RIS are changed automatically with the control of the FPGA.

\subsection{Implementation of Transceiver Module}
\begin{figure}[!t] 
	\center{\includegraphics[width=0.75\linewidth]{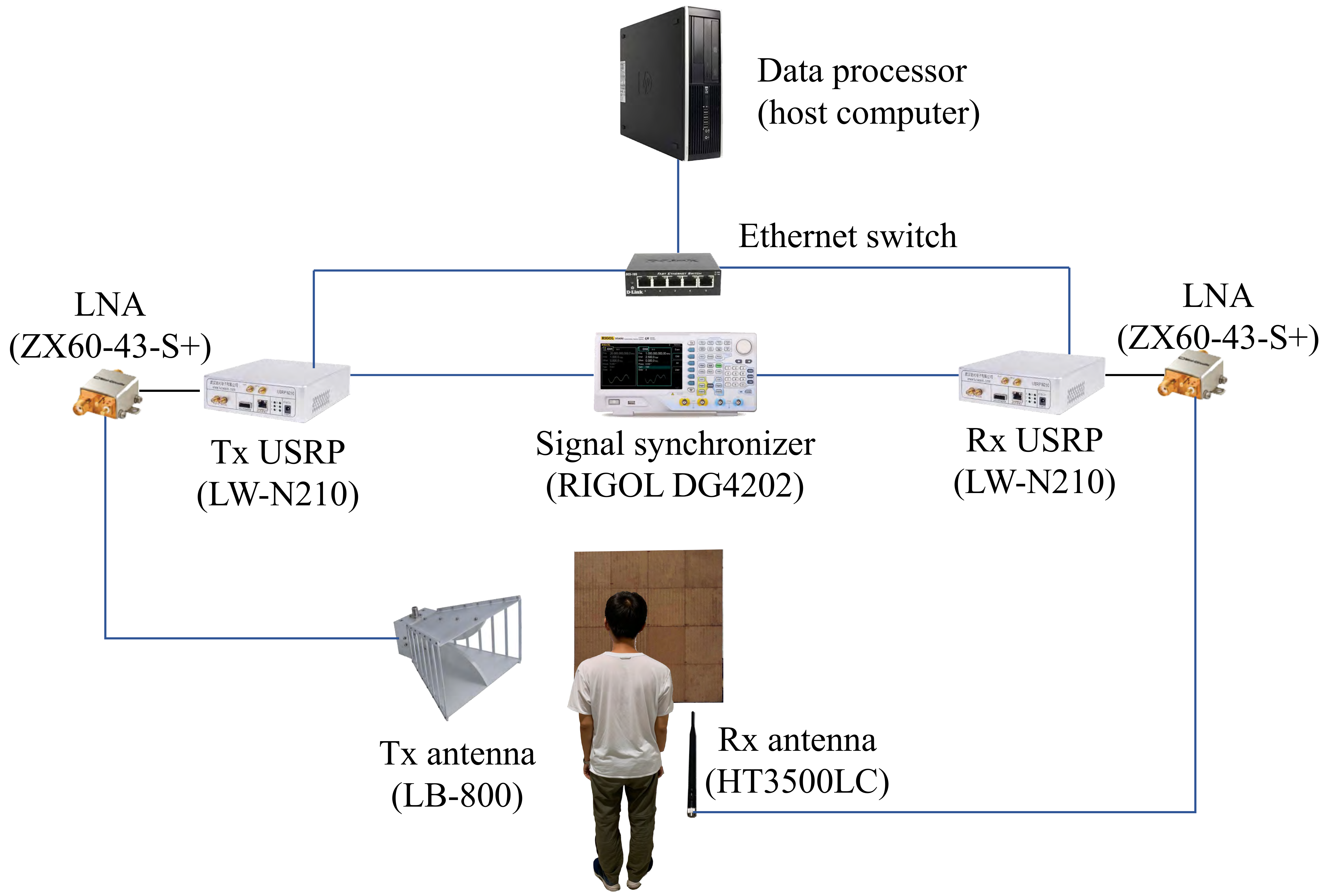}}
	\vspace{-1.em}
	\caption{
	Components of the transceiver module.}
	\label{fig: transceiver}
\end{figure}

As shown in Fig.~\ref{fig: transceiver}, the transceiver module of the designed RIS-based posture recognition system consists of the following components:
\begin{itemize}
\item \textbf{Tx and Rx USRP devices}: We implement the transmitter and receiver based on two~USRPs~(LW-N210), i.e., a Tx USRP and an Rx USRP, which are capable of converting baseband signals to RF signals and vise versa. 
The USRP is composed of the hardwares such as the RF modulation/demodulation circuits and baseband processing unit and can be controlled using software~\cite{Holland2015Universal}.
\item \textbf{Low-noise amplifiers~(LNAs)}: Two LNA~(ZX60-43-S+) are connected to the input port and output port of the two USRP, respectively, which amplify the transmitted and received RF signals of the USRPs. The LNAs connect the Tx and Rx USRP with the Tx and Rx antennas, respectively.
\item \textbf{Tx and Rx antennas}: The Tx antenna is a directional double-ridged horn antenna~(LB-800), and the Rx antenna an omnidirectional vertical antenna~(HT3500LC). Both antennas are linearly polarized.
\item \textbf{Signal synchronizer}: For the Rx USRP to obtain the relative phases and amplitudes of the received signals with respect to the transmitted signals of the Tx USRP, we employ a signal source~(RIGOL DG4202) to synchronize the Tx and Rx USRPs. The signal source provides the reference clock signal and the pulses-per-second~(PPS) signal to the USRPs, which ensures the modulation and demodulation of the USRPs to be coherent. 
\item \textbf{Ethernet switch}: The Ethernet switch connects the USRPs and a host computer to a common Ethernet, where they exchange the transmission and received signals.
\item \textbf{Data processor}: The data processor is a host computer which controls the two USRPs by using Python programs based on GNU packet~\cite{blossom2004gnu}. Besides, the host computer extracts the measurement vectors from the received signals of the Rx USRP and handles the them by the NN trained by Algorithm~\ref{alg: back-propagation algorithm} to obtain the decisions of postures.
\end{itemize}

\ifx\allfiles\undefined
\end{document}
\fi
\fi
\ifBreakPage
\newpage
\fi

\ifShowRes
\ifx\allfiles\undefined
\documentclass[onecolumn,journal,draftclsnofoot,12pt]{IEEEtran}

\usepackage{bm}
\usepackage{amssymb}
\usepackage{amsmath}
\usepackage{amsfonts}
\usepackage{graphicx}
\usepackage{subfigure}
\usepackage{cite}
\usepackage{enumerate}
\usepackage{url}
\usepackage{epstopdf}
\usepackage{float}
\usepackage{color}
\usepackage{multirow}

\usepackage[normalem]{ulem}  
\newcommand{\beq}{\begin{equation}}
\newcommand{\eeq}{\end{equation}}
\newcommand{\blue}{\textcolor[rgb]{0,0,0.7}}
\newcommand{\pasted}{\textcolor[rgb]{0.3,0,0}}
\allowdisplaybreaks[4]
\makeatletter  
\newif\if@restonecol  
\makeatother  
\let\algorithm\relax  
\let\endalgorithm\relax  
\usepackage[linesnumbered, lined, commentsnumbered, ruled]{algorithm2e}
\usepackage{amsmath}  

\usepackage{makecell} 

\renewcommand{\Pr}{\mathrm{Pr}}
\newcommand{\diag}{\mathrm{diag}}
\newcommand{\rank}{\mathrm{rank}}
\newcommand{\tx}{\mathrm{Tx}}
\newcommand{\rx}{\mathrm{Rx}}
\newcommand{\rl}{\mathrm{rl}}
\newcommand{\mau}{\mathrm{mu}}
\newcommand{\sign}{\mathrm{sign}}

\DeclareMathOperator*{\argmin}{argmin}
\DeclareMathOperator*{\argmax}{argmax}

\newif\ifshowannote
\showannotetrue

\newtheorem{definition}{\bf Definition}
\newtheorem{proposition}{\bf Proposition}
\newtheorem{theorem}{\bf Theorem}
\newtheorem{lemma}{\bf Lemma}
\newtheorem{proof}{Proof}

\begin{document}
\fi
\section{Simulation and Experimental Results}
\label{sec: results}

In this section, we first describe the system setting for the simulation and experiments and then specify the adopted parameters.
Simulation results and experimental results are then provided and discussed.

\subsection{System Setting for Simulation and Experiment}
\begin{figure}[!t] 
	\center{\includegraphics[width=0.9\linewidth]{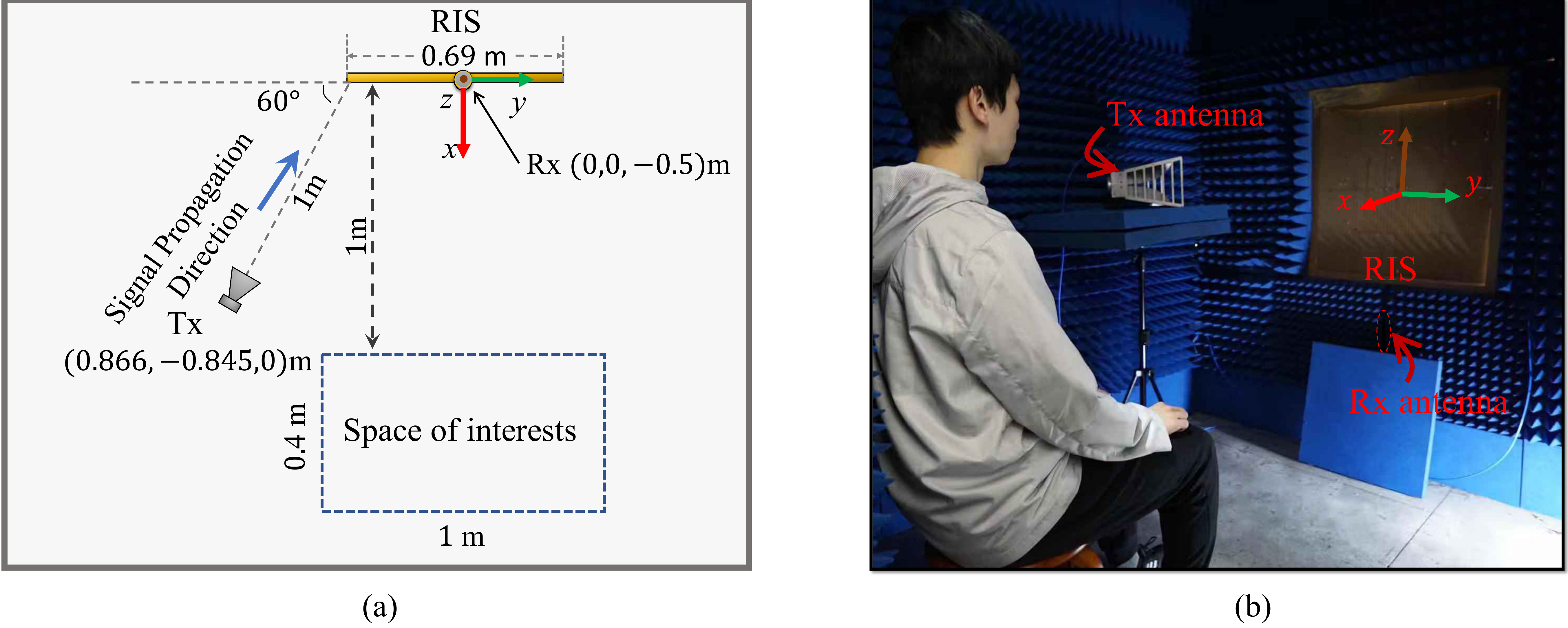}}
	\caption{Layout of the simulation and experiment of the RIS-based posture recognition system. (a) is the layout used for simulation, and (b) is the layout for the experiment, which is arranged in accordance with the layout in (a).}
	\label{fig: simul and exper envir setting}
\end{figure}
In the simulation, the layout of the RIS, the Tx and Rx antennas, and the space of interest is depicted as Fig.~\ref{fig: simul and exper envir setting}~(a).
To be specific, the origin of the 3D-coordinate is located at the center of the RIS, and the RIS is in the $y-z$ plane.
Besides, the $z$-axis is vertical to the ground and pointing upwards, and the $x$- and $y$-axes are parallel to the ground.

The Tx antenna is located more than $10\lambda$ (around $0.936$~m) away from the RIS, so that the incidence signals on the RIS elements can be approximated as a plane wave.
Therefore, the incidence angles of the incidence signals on the RIS elements are approximately the same, which is $(60^\circ, 0^\circ)$.
The human body is in the space of interest, which is a cuboid region located at $1$~m from the RIS.
Since the space of interest is behind the Tx antenna and the Tx antenna is directional horn antenna, no LoS signal path from the Tx antenna to the space of interest exists.
The side lengths of the space of interest are $l_x = 0.4$~m, $l_y = 1.0$~m, and $l_z = 1.6$~m.
Besides, the space of interest is further divided into $M = 80$ cubics with side length $0.2$~m.

Besides, we obtain the reflection coefficient function of the RIS elements by using the CST.
We model an RIS element according to Fig.~\ref{fig: RIS} and simulate the far-field radiation pattern under the stimulation of a $z$-axis polarized plane wave with the incidence angle $(60^\circ, 0^\circ)$.
The simulation is executed in the frequency domain with the unit-cell boundary.
Moreover, we project the results onto the $z$-axis to obtain the reflection coefficients since the Tx and Rx antenna are both linearly-polarized along the $z$-axis.

Fig.~\ref{fig: simul and exper envir setting}~(b) shows the experiment environment and layout of the implemented system.
We conduct practical experiments on the implemented RIS-based posture recognition system in a low-reflection environment, where the walls are covered with the wave-absorbing materials. 
This environment set-up imitates a vast and empty room, where the RIS is fixed on a wall with a low reflection rate for the $3.2$~GHz wireless signals.
The Tx and Rx antennas are in the room, while the USRPs, host computer, and the connecting circuits are located behind the RIS.

In Table~\ref{table: simul and experi parameters}, we summarize the adopted parameters in the simulation. 
The values of the parameters are taken according to the ones that we input to the algorithms and the ones that we obtained from the specifications of the employed devices, i.e., the antennas and the LNA.

\begin{table}
\centering
	\caption{Simulation Parameters.}	\label{table: simulation parameters}
	\vspace{-1em}
\includegraphics[width=0.9\linewidth]{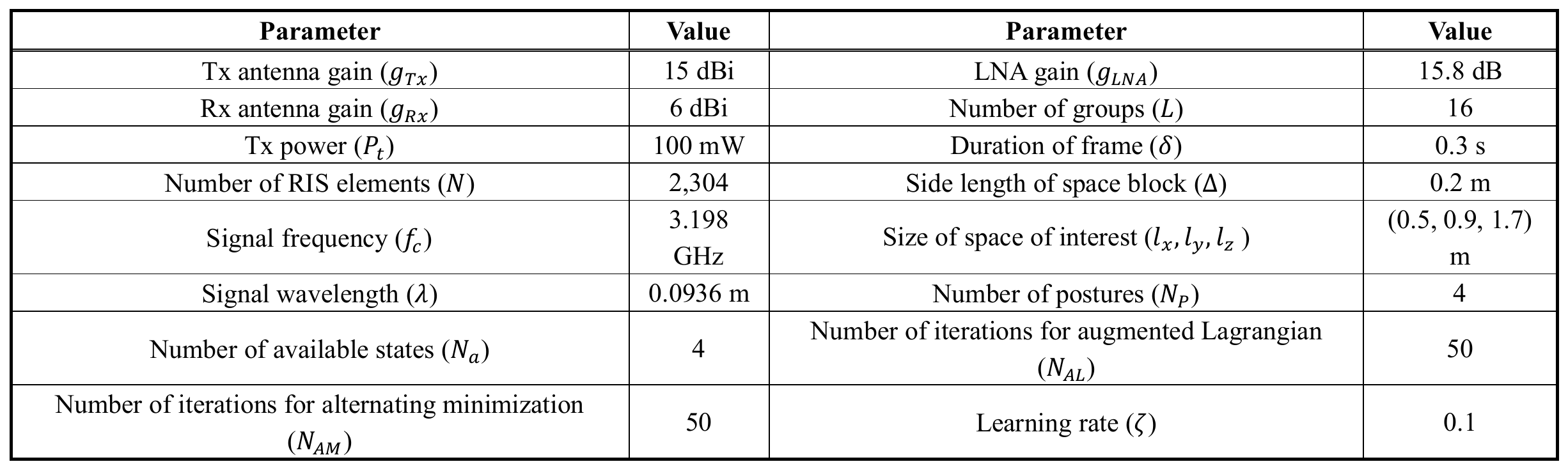}
\label{table: simul and experi parameters}
\end{table}


\subsection{Simulation Results}

\begin{figure}[!t] 
	\center{\includegraphics[width=0.55\linewidth]{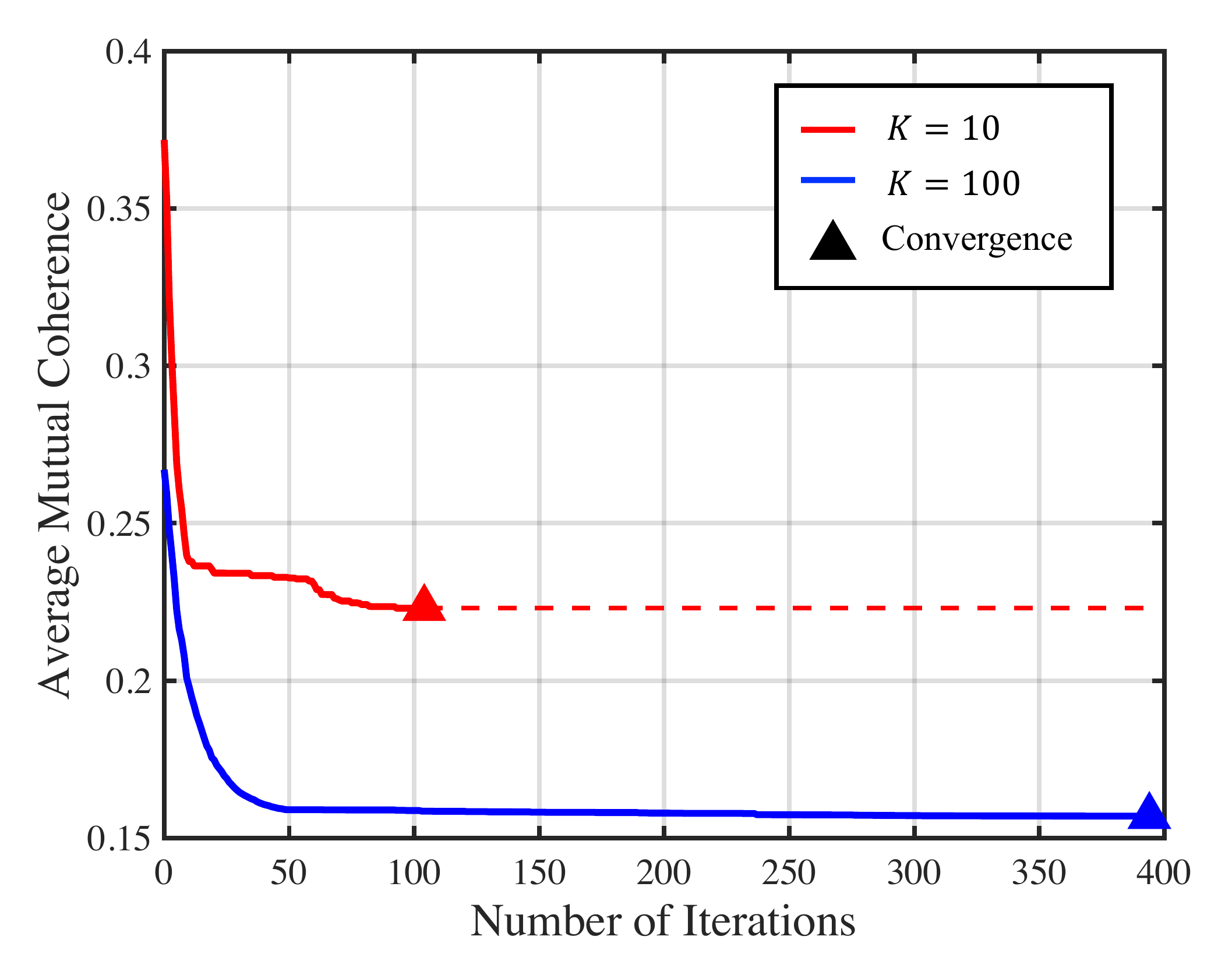}}
	\vspace{-1.5em}
	\caption{Average mutual coherence of measurement matrix vs. the number of iterations in FCAO algorithm, under different numbers of frames.}
	\label{fig: mutual coherence curve}
\end{figure}

Fig.~\ref{fig: mutual coherence curve} shows the average mutual coherence of $\bm \Gamma$ vs. the number of iterations in Algorithm~\ref{alg: ao algorithm}, under different numbers of frames $K$.
It can be observed that average mutual coherence decreases with the number of iterations, which verifies the effectiveness of the FCAO algorithm.
Besides, it can also be seen that the converged optimal average mutual coherence of $\bm \Gamma$ decreases with number of frames $K$.
This can be explained as follows.
Since $\bm \gamma_m$, i.e., the $m$-th column of $\bm \Gamma$ indicates the measurement of the $m$-th space block.
Since $\bm \gamma_m = \bm T\bm \alpha_m$~($\forall m\in[1,M]$), to reduce the mutual coherence of $\bm \Gamma$, it requires that the configuration matrix $\bm T$ maps $\bm \alpha_m$ to $\bm \gamma_m$ where different $\bm \gamma_m$ have large elements at different dimensions.
When $K$ is large, the number of dimensions of $\bm \gamma_m$~($\forall m\in[1,M]$) is large, and therefore, the probability of finding $\bm T$ to distribute the large components of $\bm \gamma_m$ at different dimensions is larger.

\begin{figure}[!t] 
	\center{\includegraphics[width=0.73\linewidth]{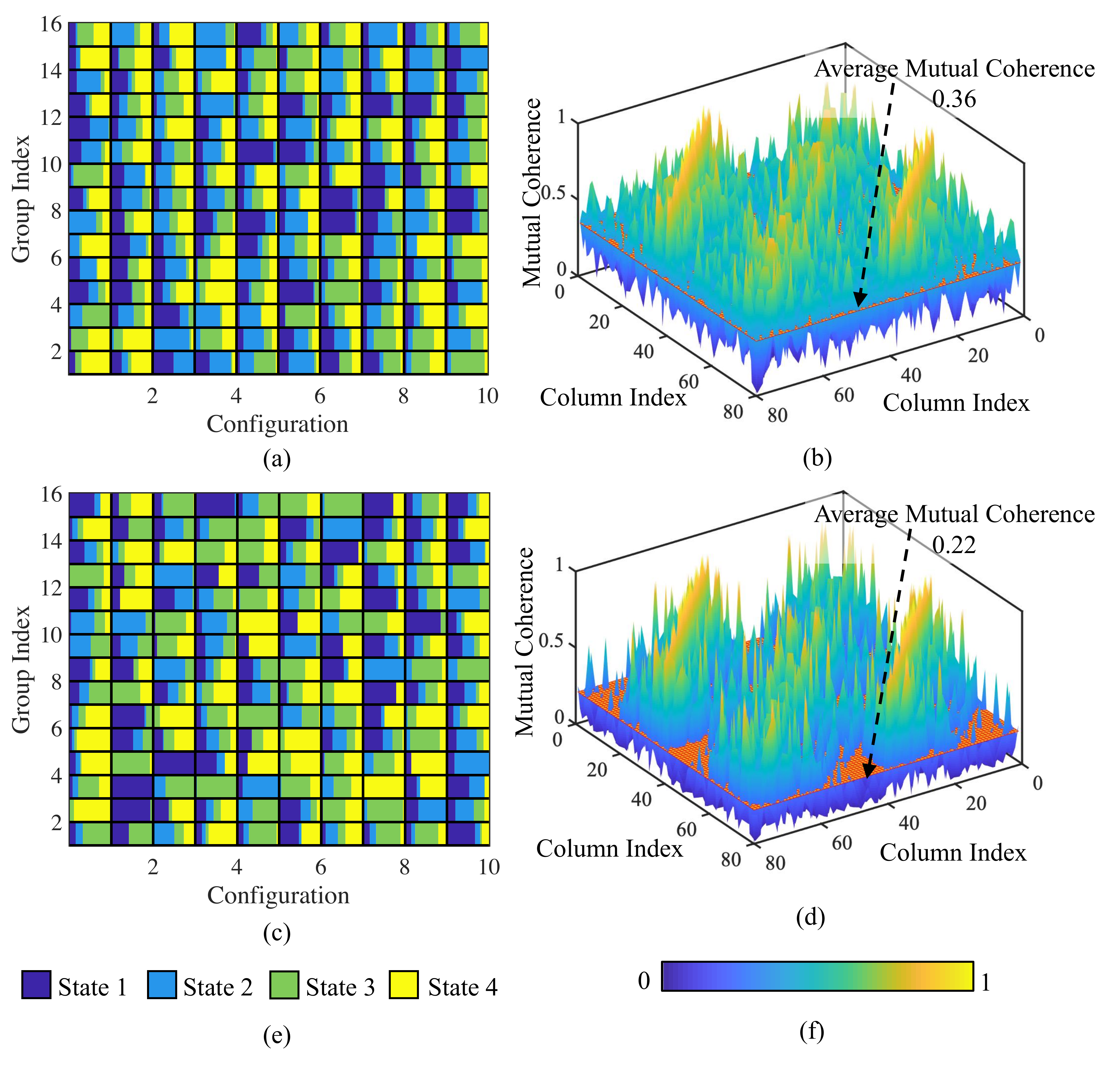}}
	\vspace{-1em}
	\caption{(a) Illustration of a random configuration; (b) Mutual coherence of the measurement matrix given the random configuration; (c) Illustration of the optimized configuration obtained by Algorithm~\ref{alg: ao algorithm}; (d) Mutual coherence of the measurement matrix given the optimized configuration; (e) Colors that represents the duration of different states; (f) Color scaling representing the coherence to be from $0$ to $1$.}
	\label{fig: config sequence compare}
\end{figure}

Besides, we illustrate the configuration matrix and the corresponding coherence vectors before and after the configuration matrix optimization with $K=10$.
The configuration sequence shown in Fig.~\ref{fig: config sequence compare}~(a) is a random configuration matrix where the duration of all the states in each configuration follows the same uniform distribution with a fixed sum equaling to $\delta$.
Fig.~\ref{fig: config sequence compare}~(c) shows the coherence of column vectors of $\bm\Gamma$, i.e., $u_{m,m'}$~($\forall m,m'\in[1,M],~m\neq m'$).
The configuration matrix in Fig.~\ref{fig: config sequence compare}~(c) is the optimized configuration matrix obtained by using Algorithm~\ref{alg: ao algorithm}, and Fig.~\ref{fig: config sequence compare}~(d) shows the mutual coherence of the measurement matrix corresponding to it.
Comparing Figs.~\ref{fig: config sequence compare}~(b) and~(d), it can be seen that most of the coherence values corresponding to the optimized configuration matrix is lower than that corresponding to the random configuration matrix.
Besides, it can also be observed that the average mutual coherence corresponding to the optimized configuration matrix is significantly reduced.
Therefore, the effectiveness of the FCAO algorithm is verified.

\begin{figure}[!t] 
	\center{\includegraphics[width=0.6\linewidth]{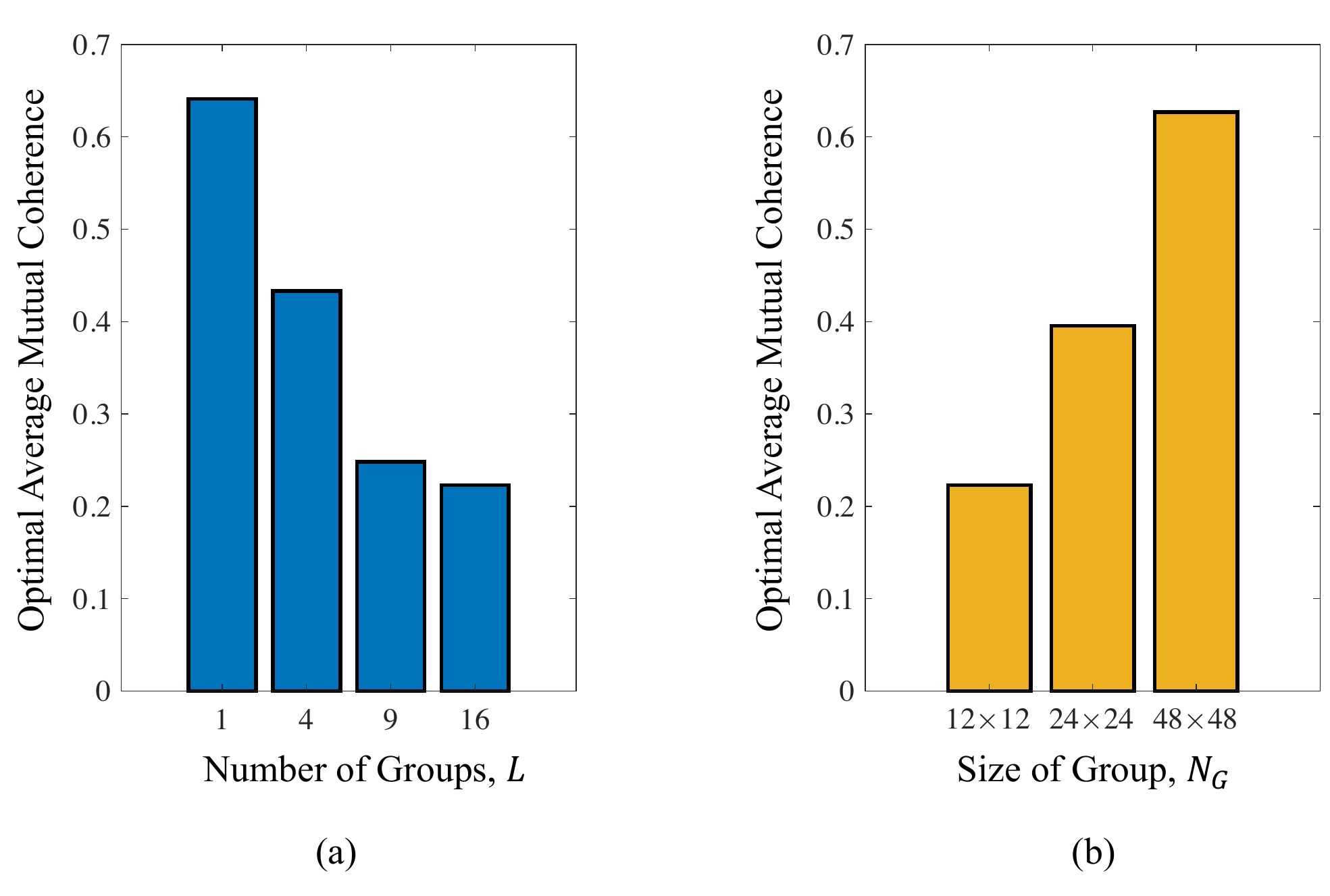}}
	\vspace{-1.5em}
	\caption{Optimal average mutual coherence values vs.~(a) number of groups, $L$, given $N_G = 12\times 12$; and~(b) size of group, $N_G$, given number of elements $N = 48\times 48$.}
	\label{fig: coherence group}
\end{figure}
Figs.~\ref{fig: coherence group}~(a) and~(b) show the optimal average mutual coherence obtained by Algorithm~\ref{alg: ao algorithm} under different sizes of the RIS and different sizes of group, respectively.
In Fig.~\ref{fig: coherence group}~(a), the size of the RIS is determined by the number of groups, $L$, and each group contains $N_G = 12\times 12$ RIS elements. 
It can be seen that the optimal average mutual coherence decreases with the number of groups that the RIS contains, i.e., the size of the RIS.
In Fig.~\ref{fig: coherence group}~(b), the size of the RIS is fixed, and the RIS contains $N=48\times 48$ elements.
The value of $N_G$ determines the number of groups of the RIS, which can be controlled independently.
It can be seen that the optimal average mutual coherence increases with the size of groups.

Based on Section~\ref{sec: config. design opt.}, the average mutual coherence is negatively related to the recognition accuracy, which is proved in the experimental results in the following subsection.
Therefore, the results shown in Figs.~\ref{fig: coherence group}~(a) and~(b) also indicate that the posture recognition accuracy of the system increase with the size of the RIS and the number of independently controllable groups.

\subsection{Experimental Results}
\begin{figure}[!t] 
	\center{\includegraphics[width=0.8\linewidth]{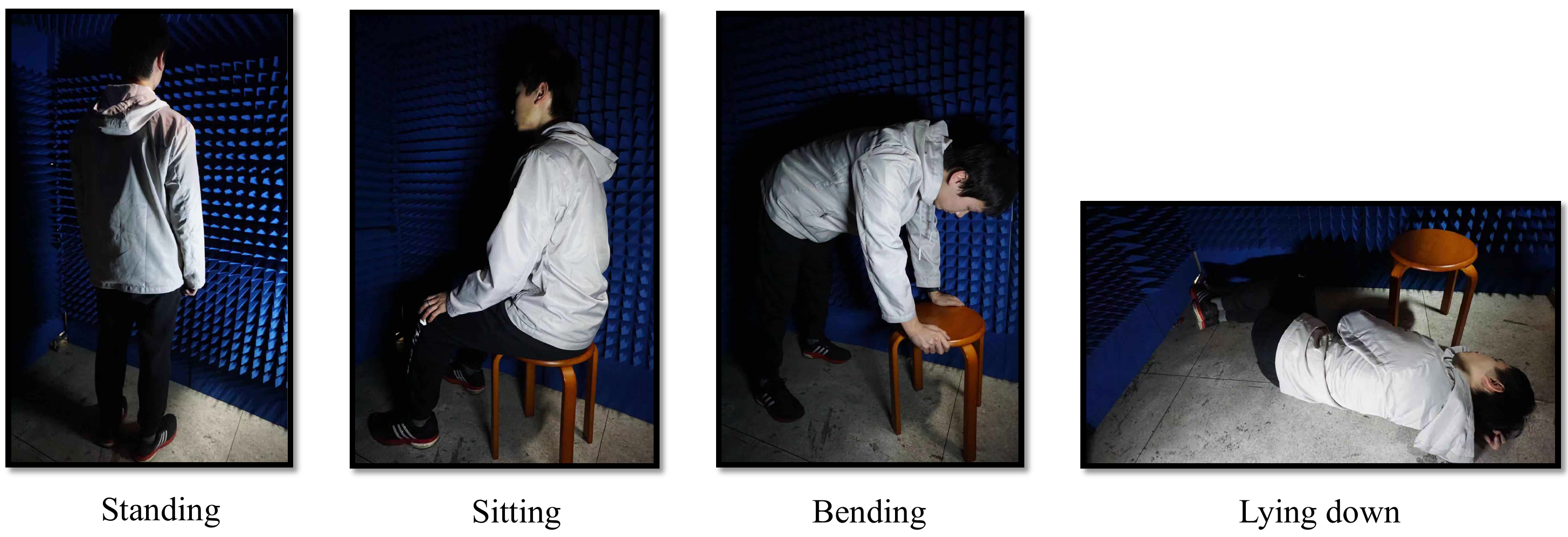}}
	\vspace{-1em}
	\caption{Human postures for recognition: standing, sitting, bending, and lying down.}
	\label{fig: human postures}
\end{figure}
We use the implemented posture recognition system described in Section~\ref{sec: system implementation} to perform the experiments.
We set $K=10$, i.e., there are $10$ frames in each recognition period.
As for the human postures for recognition, we consider the $N_P = 4$ postures shown in Fig.~\ref{fig: human postures}.
For each posture, we collect $150$ labeled measurement vectors in the random configuration matrix case and the optimized configuration matrix case, respectively, and form the data sets.
Then, we divide the data set into the training set and the testing set in each case.
The training data set contains $120$ labeled measurement vectors, and the testing data set contains $30$ measurement vectors to be processed.
In each case, we train the decision function, i.e., the NN using the training data set based on Algorithm~\ref{alg: back-propagation algorithm}.
Then, we use the trained NN to process the measurement vectors in the testing set and record the probabilities of the decisions on each posture.

For comparison, we also perform the same experiments in the \emph{non-configurable environment} case, which serves as a benchmark.
In the non-configurable environment case, the RIS elements are fixed to state $\hat{s}_1$, and therefore, the system work as a single-antenna RF sensing system.

\begin{figure}[!t] 
	\center{\includegraphics[width=0.55\linewidth]{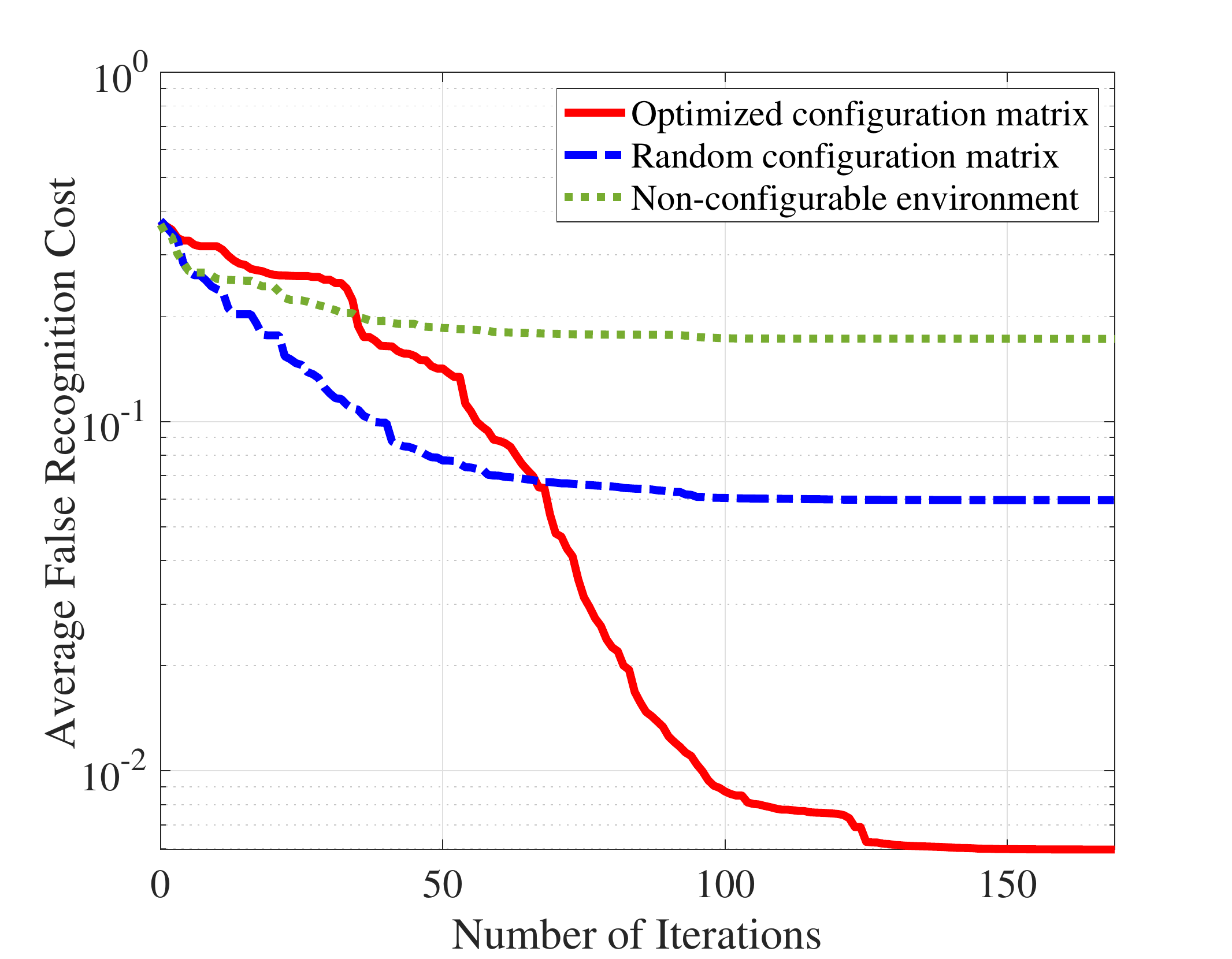}}
	\caption{Average false recognition cost vs. the number of iterations of Algorithm~\ref{alg: back-propagation algorithm} given optimized configuration matrix and random configuration matrix. $K = 10$.}
	\label{fig: posture recognition cost}
\end{figure}
Fig.~\ref{fig: posture recognition cost} shows the average false recognition cost vs. the number of iterations of Algorithm~\ref{alg: back-propagation algorithm}, where the costs of true and false recognition are set to $0$ and $1$, respectively, i.e., $\chi_{i,i'} = 0$ if $i=i'$; otherwise, $\chi_{i,i'} = 1$.
It can be observed that the average false recognition cost decreases with the number of iterations.
Besides, the converged value of average false recognition cost using the optimized configuration matrix is about $10$ times smaller than that using a random configuration matrix.
This verifies that the optimized configuration matrix, which has a measurement matrix with low average mutual coherence, can results in lower false recognition cost compared to the random configuration matrix.
Moreover, by comparing with the benchmark case, we can observe that the capability of the RIS to customize the environment helps the RF sensing system to reduce the average false recognition cost.

\begin{figure}[!t] 
	\center{\includegraphics[width=1\linewidth]{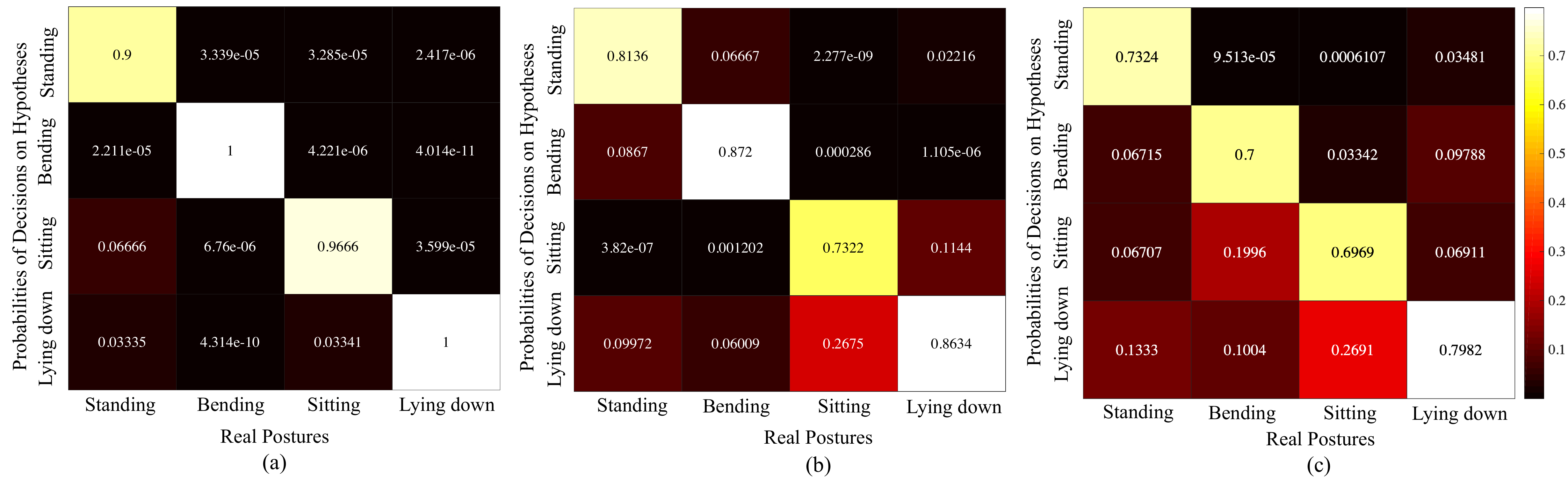}}
	\vspace{-2em}
	\caption{Posture recognition accuracy in the (a) optimized configuration matrix case; (b) random configuration matrix case; and (c) non-configurable environment case. $K = 10$.}
	\label{fig: posture recognition precision}
\end{figure}
Figs.~\ref{fig: posture recognition precision}~(a) and~(b) shows the accuracy of the posture recognition in the optimized configuration matrix case and the random configuration matrix case, respectively.
It can be seen that in the optimized configuration matrix case, the recognition accuracy is much higher than that in the random configuration matrix case.
This verifies that the optimized configuration matrix, which has a measurement matrix with low average mutual coherence, can result in higher recognition accuracy in the practical posture recognition system.
Besides, it can be observed that the system with optimized configuration can achieve $14.6\%$ higher recognition accuracy compared with that with random configuration.
Moreover, by comparing with the benchmark case, we can observe that the capability of the RIS to customize the environment increases the posture recognition accuracy of RF sensing systems with $23.5\%$.

\ifx\allfiles\undefined
\end{document}
\fi
\fi
\ifBreakPage
\newpage
\fi

\section{Conclusion}
\label{sec: conclusion}
In this paper, we have designed an RIS-based posture recognition system.
To facilitate the configuration design, we have proposed a frame-based periodic configuring protocol.
Based on the protocol, we have formulated the optimization problem for false recognition cost minimization.
To solve the problem, we have decomposed it into the configuration matrix and the decision function optimization problems, and proposed the FCAO algorithm and the supervised learning algorithm to solve them, respectively.
Besides, based on USRPs, we have implemented the designed system and executed posture recognition experiments in practical environments.
Simulations have verified that the FCAO algorithm can obtain the optimal configuration matrix, which leads to a measurement matrix with low average mutual coherence.
The experimental results prove that the configuration matrix with lower average mutual coherence has higher recognition accuracy and a lower false recognition cost.
Besides, combing the simulation and experimental results, we have shown that the posture recognition accuracy increase with the size of the RIS and the number of independently controllable groups.  
Moreover, compared with the random configuration and the non-configurable environment cases, the optimized configuration can achieve $14.6\%$ and $23.5\%$ higher recognition accuracy, respectively.
\begin{appendices}

\ifx\allfiles\undefined
\documentclass[onecolumn,journal,draftclsnofoot,12pt]{IEEEtran}

\usepackage{bm}
\usepackage{amssymb}
\usepackage{amsmath}
\usepackage{amsfonts}
\usepackage{graphicx}
\usepackage{subfigure}
\usepackage{cite}
\usepackage{enumerate}
\usepackage{url}
\usepackage{epstopdf}
\usepackage{float}
\usepackage{color}
\usepackage{multirow}

\usepackage[normalem]{ulem}  
\newcommand{\beq}{\begin{equation}}
\newcommand{\eeq}{\end{equation}}
\newcommand{\blue}{\textcolor[rgb]{0,0,0.7}}
\newcommand{\pasted}{\textcolor[rgb]{0.3,0,0}}
\allowdisplaybreaks[4]
\makeatletter  
\newif\if@restonecol  
\makeatother  
\let\algorithm\relax  
\let\endalgorithm\relax  
\usepackage[linesnumbered, lined, commentsnumbered, ruled]{algorithm2e}
\usepackage{amsmath}  

\usepackage{makecell} 

\renewcommand{\Pr}{\mathrm{Pr}}
\newcommand{\diag}{\mathrm{diag}}
\newcommand{\rank}{\mathrm{rank}}
\newcommand{\tx}{\mathrm{Tx}}
\newcommand{\rx}{\mathrm{Rx}}
\newcommand{\rl}{\mathrm{rl}}
\newcommand{\mau}{\mathrm{mu}}
\newcommand{\sign}{\mathrm{sign}}

\DeclareMathOperator*{\argmin}{argmin}
\DeclareMathOperator*{\argmax}{argmax}

\newif\ifshowannote
\showannotetrue

\newtheorem{definition}{\bf Definition}
\newtheorem{proposition}{\bf Proposition}
\newtheorem{theorem}{\bf Theorem}
\newtheorem{lemma}{\bf Lemma}
\newtheorem{proof}{Proof}

\begin{document}
\fi

\section{Augmented Lagrangian Algorithm for (P5)}
\label{appx: augmented Lagrangian alg.}

The augmented Lagrangian method finds a local optimal solution to~(P5) by minimizing a sequence of~(\ref{equ: convenient aug. Lagrangian}) where $\bm \beta$ and $\rho$ are held fixed in each iteration.
Specifically, the sequence of augmented Lagrangian minimization can be solved using an \emph{alternating minimization procedure},  in which $\bm u$ is updated while $\bm t_k$ is held fixed, and vice versa.
The alternating minimization procedure can be described as follows.

\textbf{1. Update Step for $\bm u$}:
We consider the $\bm u$ update step in the alternating minimization procedure for~(\ref{equ: convenient aug. Lagrangian}) minimization given $\bm \beta$, $\rho$ and fixed $\bm T_{-k}$, and $\bm t_{k}$.
As the first step, we introduce the auxiliary variable $z_{m,m'}= \frac{\bm \gamma_m^T \bm\gamma_{m'}}{\|\bm \gamma_m\|_2\cdot \|\bm \gamma_{m'}\|_2} - \beta_{i,j}/\rho$ and arrange the auxiliary variables into a vector $\bm z = (z_{1,2}, z_{1,3}, \dots, z_{1,M}, z_{2,3},\dots, z_{M-1,M})$.
Then, update $\bm u$ to solve the minimization for~(equ: convenient aug. Lagrangian) can be handled by the proximal gradient method~\cite{Obermeier2018Sensing}, which is equivalent to solving
\beq
\label{opt: update step for u}
\mathrm{P}_{\bm u}:~\min_{\bm u} \|\bm u\|_{1} + \frac{\rho}{2}\|\bm u - \bm z\|_2^2.                                                                                                                                                                                                                                                                                                                                                                                                                                                                                                                                                                                                                                                                                                                                                                                                                                                                                                            
\eeq
Since the sum of norm functions are convex, ($\mathrm{P}_{\bm u}$) is an unconstrained convex optimization problem, which can be solved efficiently by using existing convex optimization algorithms~\cite{Boyd_CONVEX}.   
   
\textbf{2. Update Step for $\bm t_k$}:
We then consider the $\bm t_k$ update step.
We first introduce the auxiliary variable $\kappa_{m,m'} = u_{m,m'}+\beta_{m,m'}/\rho$.
Then, the minimization for the augmented Lagrangian given fixed $\bm u$ can be reduced to the following non-convex optimization problem:
\begin{align}
(\mathrm{P}_{\bm t_k})
\min_{\bm d_k} \quad
&\sum_{1\leq m<m'\leq M} {\rho\over 2}\left|  \frac{\bm \gamma_m^T \bm\gamma_{m'}}{\|\bm \gamma_m\|_2\cdot \|\bm \gamma_{m'}\|_2} - \kappa_{m,m'}\right|^2\\
s.t. \quad
& \bm 1^T\tilde{\bm t}_{k,l} = 1,~\forall l\in[1,L],\\
& \tilde{\bm t}_{k,l} \succeq 0, ~\forall l \in [1,L].
\end{align}

Due to the complicated objective function, the optimum to $(\mathrm{P}_{\bm t_k})$ is hard to solve.
Nevertheless, an approximate update for $\bm t_k$ can be found using the proximal gradient method proposed in~\cite{Obermeier2017Sensing}.

After the alternating minimization procedure, dual variable $\beta_{m,m'}$ is updated by
\begin{equation}
\label{equ: beta update}
	\beta_{m,m'}=\beta_{m,m'}  + \rho\left(
u_{m,m'} - \frac{\bm \gamma_m^T \bm\gamma_{m'}}{\|\bm \gamma_m\|_2\cdot \|\bm \gamma_{m'}\|_2}
	\right).
\end{equation}
Besides, $\rho$ is updated by using the method described in~\cite{Migliore2011Compressed}.
In summary, the algorithm to solve the augmented Lagrangian function minimization is proposed as Algorithm~\ref{alg: aug. Lagrangian update procedure}.
\begin{algorithm}[!t]  \label{alg: aug. Lagrangian update procedure}
\small
\caption{Alternating minimization algorithm for solving (P5).}
	\SetKwInOut{Input}{Input}
	\SetKwInOut{Output}{Output}
\Input{
	Number of iterations for solving augmented Lagrangian function~($N_{\mathrm{AL}}$);
	Number of iterations for the alternating minimization procedure~($N_{\mathrm{AM}}$);
	Maximum iteration number $N_{\max}$;
	$\bm t_{k}^{(0)}$, $\bm u^{(0)}$. 
}
\Output{
Optimized $\bm d_{k}^*$ and $\bm u^*$ which are a sub-optimal solution for (P5).
}


Calculate $u_{m,m'}^{(0)} = \frac{\bm (\gamma_m^{(0)})^T \bm\gamma_{m'}^{(0)}}{\|\bm \gamma_m^{(0)}\|_2\cdot \|\bm \gamma_{m'}^{(0)}\|_2}$ based on $\bm t_k^{(0)}$, and set random $\beta_{m,m'}^{(1)}\in(0,1)$, $\forall m,m'\in[1,M],~m<m'$\;

\For{$ a= 1,\dots, N_{\mathrm{AL}}$}
{
	Set $\bm u^{(a, 0)} = \bm u^{(a-1)}$, $\bm t^{(a,0)}_k = \bm t^{(a-1)}_k$\;
	
	\For{$b = 1,\dots, N_{\mathrm{AD}}$}
	{
		\textbf{Update $\bm u$}: Obtain $\bm u^{(a,b+1)}$ by solving~$\mathrm{P}_{\bm u}$ using convex optimization algorithm in~\cite{Boyd_CONVEX}, given $\bm t_{k}^{(a,b)}$, $\bm \beta^{(a)}$, and $\rho^{(a)}$\;
		
		\textbf{Update $\bm t_{k}$}: Obtain $\bm t^{(a,b+1)}_k$ by solving~$\mathrm{P}_{\bm t_k}$ using proximal gradient method in~\cite{Obermeier2017Sensing}, given $\bm u^{(a,b+1)}$, $\bm \beta^{(a)}$, and $\rho^{(a)}$\;
	}
	
	Compute the dual variables $\bm\beta^{(a+1)}$ by~(\ref{equ: beta update}) and 
	compute $\rho^{(a+1)}$ by using the method in~\cite{Migliore2011Compressed}.
}
\end{algorithm}

\ifx\allfiles\undefined
\end{document}
\fi

\end{appendices}
\bibliographystyle{IEEEtran}
\bibliography{bibilio}

\end{document}